\documentclass[twocolumn,superscriptaddress]{revtex4-2}
\usepackage[colorlinks=true, citecolor=blue, urlcolor=blue, linkcolor=red]{hyperref}
\usepackage{amsmath,amssymb,tikz,scalerel,physics}
\hypersetup{colorlinks=true, linkcolor=red, citecolor=blue, urlcolor=blue}
\usetikzlibrary{svg.path}
\definecolor{orcidlogocol}{HTML}{A6CE39}
\tikzset{
	orcidlogo/.pic={
		\fill[orcidlogocol] svg{M256,128c0,70.7-57.3,128-128,128C57.3,256,0,198.7,0,128C0,57.3,57.3,0,128,0C198.7,0,256,57.3,256,128z};
		\fill[white] svg{M86.3,186.2H70.9V79.1h15.4v48.4V186.2z}
		svg{M108.9,79.1h41.6c39.6,0,57,28.3,57,53.6c0,27.5-21.5,53.6-56.8,53.6h-41.8V79.1z M124.3,172.4h24.5c34.9,0,42.9-26.5,42.9-39.7c0-21.5-13.7-39.7-43.7-39.7h-23.7V172.4z}
		svg{M88.7,56.8c0,5.5-4.5,10.1-10.1,10.1c-5.6,0-10.1-4.6-10.1-10.1c0-5.6,4.5-10.1,10.1-10.1C84.2,46.7,88.7,51.3,88.7,56.8z};}}
\newcommand\orcid[1]{\href{https://orcid.org/#1}{\mbox{\scalerel*{\begin{tikzpicture}[yscale=-1,transform shape]\pic{orcidlogo};\end{tikzpicture}}{|}}}}

\begin{document}
\title{Noise mitigation in quantum teleportation}
\author{Zi-Jian Xu}
\affiliation{School of Physical Science and Technology \& Lanzhou Center for Theoretical Physics, Lanzhou University, Lanzhou 730000, China}
\affiliation{Key Laboratory of Quantum Theory and Applications of MoE \& Key Laboratory of Theoretical Physics of Gansu Province, Lanzhou University, Lanzhou 730000, China}
\author{Jun-Hong An\orcid{0000-0002-3475-0729}}
\email{anjhong@lzu.edu.cn}
\affiliation{School of Physical Science and Technology \& Lanzhou Center for Theoretical Physics, Lanzhou University, Lanzhou 730000, China}
\affiliation{Key Laboratory of Quantum Theory and Applications of MoE \& Key Laboratory of Theoretical Physics of Gansu Province, Lanzhou University, Lanzhou 730000, China}

\begin{abstract}
Permitting the transmission of unknown quantum states over long distances by using entanglement, quantum teleportation serves as an important building block for many quantum technologies. However, in the noisy intermediate-scale quantum era, the practical realization of quantum teleportation is inevitably challenged by the noise-induced decoherence. We here propose a noise-mitigation mechanism applicable in both the discrete- and continuous-variable quantum teleportation schemes. Via investigating the non-Markovian decoherence dynamics of the two types of quantum teleportation schemes, we find that, as long as a bound state is formed in the energy spectrum of the total system consisting of the involved subsystems and their respective reservoirs, the quantum superiority of the fidelity is persistently recovered. Supplying an insightful understanding of the noise-mitigation protocols, our result paves the way to the practical realization of noise-tolerant quantum teleportation.
\end{abstract}
\maketitle

\section{Introduction}
Quantum teleportation allows the transfer of an arbitrary unknown quantum state between distant parties by using quantum entanglement \cite{PhysRevLett.70.1895,Pirandola2015,HuXM2023}. It is not only a basic protocol of quantum communication, but also a building block in the realization of various large-scale quantum technologies. It plays important roles in quantum repeaters \cite{PhysRevLett.81.5932}, which are pivotal for quantum communication over long distances \cite{9023997}, quantum computation \cite{BRASSARD199843,Gottesman1999,PhysRevLett.86.5188,PhysRevLett.101.240501}, quantum networks \cite{Kimble2008}, and quantum
secret sharing \cite{PhysRevLett.124.060501}. It is also used as a tool for exploring fundamental physics, such as closed timelike curves \cite{PhysRevLett.106.040403} and black-hole evaporation \cite{Lloyd2014}. Following two parallel directions to encode information in discrete-variable \cite{PhysRevLett.70.1895} or continuous-variable \cite{PhysRevLett.80.869} systems, quantum teleportation has been realized in various systems \cite{Bouwmeester1997,ma2012quantum,PhysRevLett.123.070505,doi:10.1126/science.282.5389.706,doi:10.1126/science.1201034,Georgescu2022}. Outstanding performances have been achieved in terms of teleportation distance in aid of satellites \cite{ren2017ground}.

Being in the noisy intermediate-scale quantum era \cite{Preskill2018quantumcomputingin,Chen2023,https://doi.org/10.1002/qute.202300218}, we are generally forced to make a persistent effort to beat the unwanted decoherence effect induced by different kinds of noises in quantum technologies. An ideal teleportation process requires a noiseless quantum channel established by sharing a pure maximally entangled state. The protection of quantum channels from decoherence-induced disentanglement is a prerequisite for the realization and application of quantum teleportation \cite{PhysRevLett.76.722,lee2000entanglement,oh2002fidelity,jung2008greenberger,rao2008teleportation,yeo2009effects,PhysRevLett.87.267901,PhysRevA.85.020301,PhysRevA.90.042332,PhysRevA.92.012338,Guo2020,PhysRevResearch.3.033119,Im2021,doi:10.1073/pnas.2026250118,Harraz2022,SEIDA2022128115,Zhao2023,Roszak2023purifying,Hoke2023}. It has been found that the quantum advantages of teleportation typically disappear in a noisy setting, with decoherence degrading quantum teleportation to classical teleportation. Many strategies, for example, error correction \cite{doi:10.1073/pnas.2026250118}, weak measurement \cite{Harraz2022}, entanglement purification \cite{PhysRevLett.76.722}, and noiseless linear amplification \cite{Zhao2023}, have been proposed to overcome the destructive impacts of the decoherence on quantum teleportation. However, the analysis and methodologies considered thus far to address the decoherence effect on quantum teleportation are based on the Born-Markovian approximation. Given the inherent non-Markovian nature of the decoherence dynamics \cite{Rivas_2014,RevModPhys.88.021002,LI20181,doi:10.1080/23746149.2020.1870559}, it is expected that such treatments are insufficient. Furthermore, with the rapid development of experimental techniques, more and more quantum systems have exhibited the non-Markovian effect in experiments \cite{White2020,PhysRevA.104.022432,PhysRevLett.126.230401,yu2018experimental,lu2020observing,liu2017quantum,krinner2018spontaneous,Kwon2022}. It was really found that the non-Markovian effect may play an active role in retrieving the quantum superiority from the noises in several protocols of quantum technologies \cite{PhysRevLett.123.040402,PhysRevA.102.012217,PhysRevLett.124.140502,Taranto2021,PhysRevLett.131.050801,PhysRevLett.132.090401,PhysRevA.82.032340}. A natural question is: What is the non-Markovian effect on the noisy quantum teleportation? Although Refs. \cite{PhysRevA.108.062406,PhysRevA.102.062208} have revealed that the non-Markovian effect plays a constructive role in slowing down the the degradation of the average fidelity in quantum teleportation, its average fidelity in the long-time condition still becomes smaller than the classical limit. How to preserve the quantum superiority during the whole evolution process is still an open question.

\begin{figure*}[tbp]
\centering
\includegraphics[width=.8\columnwidth]{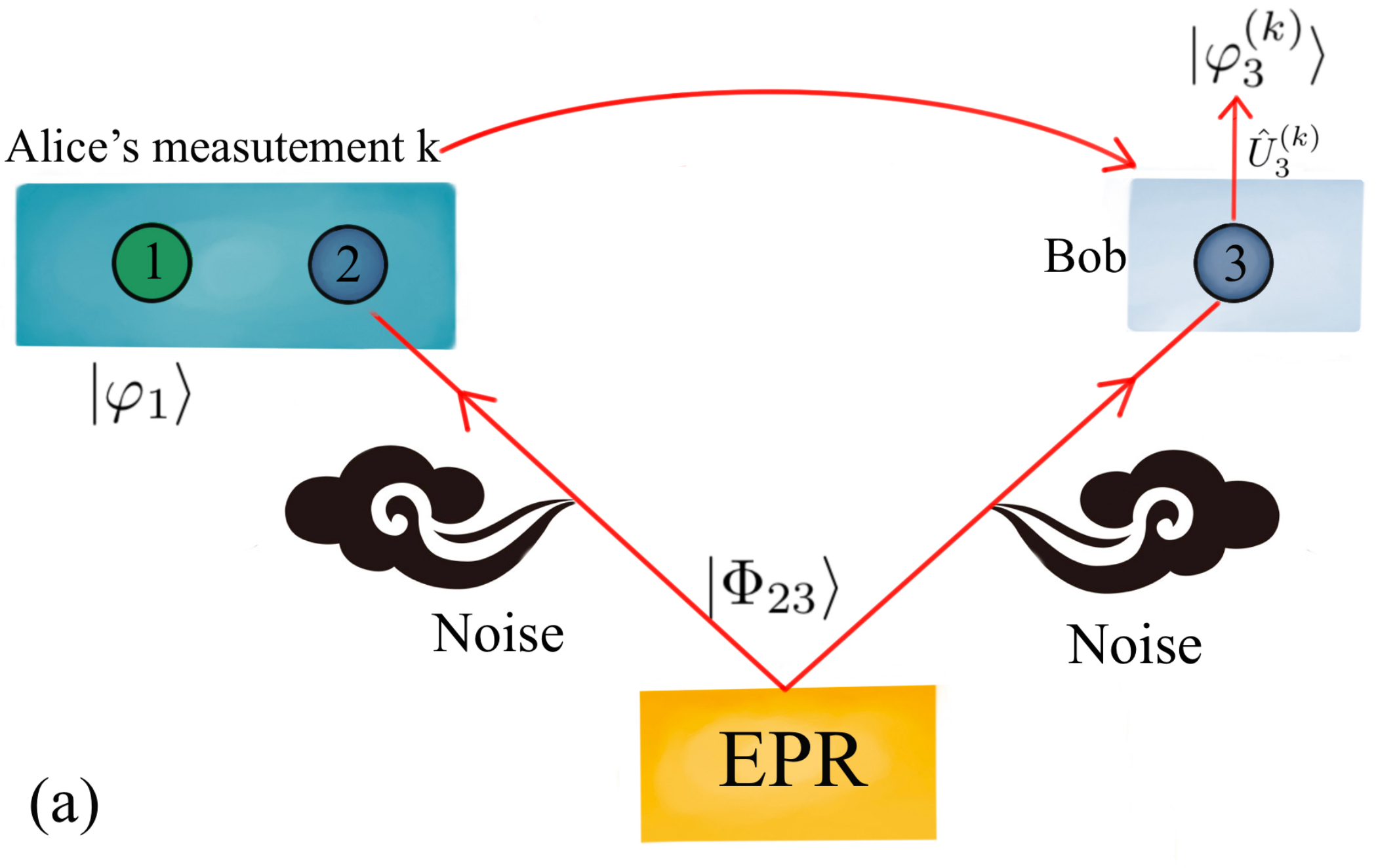}~~~~~\includegraphics[width=.8\columnwidth]{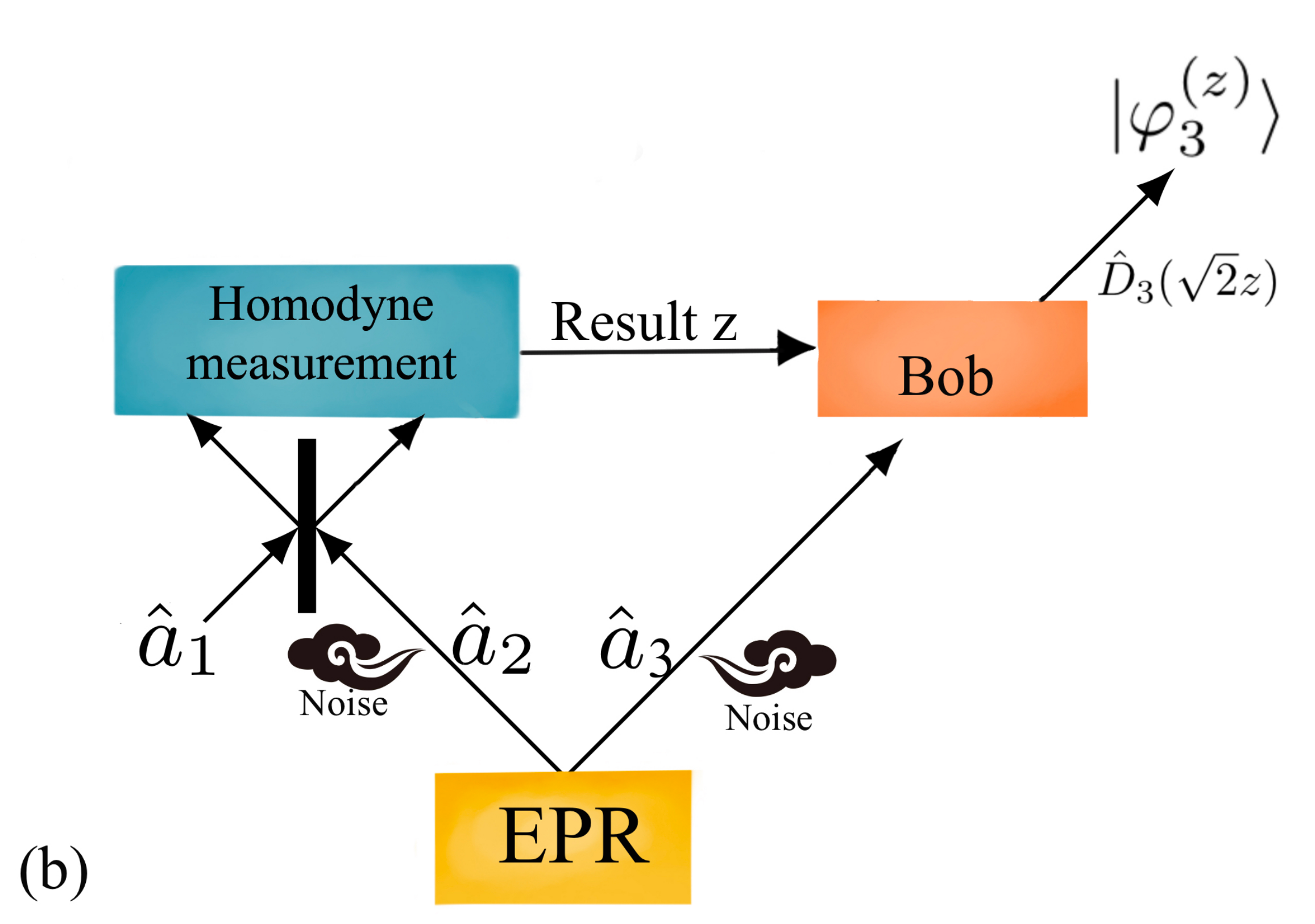}
\caption{Schematic diagrams of (a) discrete-variable and (b) continuous-variable quantum teleportation.}\label{flow diagram}
\end{figure*}

In this work, we propose a mechanism to recover the quantum superiority of teleportation influenced by the local dissipative noises. Via investigating the non-Markovian decoherence dynamics of the teleportation schemes for both the discrete- and continuous-variable systems, we find that the teleportation performance is sensitively determined by the feature of the energy spectrum of the total system formed by the involved subsystems of the quantum channels and their local reservoirs. When a bound state is present in the energy spectrum, the fidelity for both the teleportation schemes in the long-time steady state surpasses their corresponding classical limits, which are also the achievable bounds under the Born-Markovian approximate description to the decoherence. Our result demonstrates that the non-Markovian effect has a self-healing ability to rescue the teleportation from the noise-induced degradation. Refreshing our understanding of the noise effect on quantum teleportation, our result paves the way for overcoming the destructive influence of the decoherence on quantum teleportation in practical noises.

This paper is organized as follows. In Sec. \ref{Ideascm}, we begin by presenting the ideal quantum teleportation schemes for both the discrete- and continuous-variable systems. In Sec. \ref{noiseff}, we study the non-Markovian decoherence dynamics of the two teleportation schemes. The dominate role of the bound state of the involved subsystems of the quantum channels and their respective local reservoirs in the teleportation performance is revealed in this section. Section \ref{numres} is devoted to the numerical calculation to verify our result. A physical system to verify our result is suggested in Sec. \ref{phdrd}. We conclude the paper in Sec. \ref{Dsccon}.

\section{Ideal quantum teleportation}\label{Ideascm}
Quantum teleportation is the transfer of an unknown quantum state over a long distance by using a quantum channel formed by entanglement and a classical channel via classical communication \cite{PhysRevLett.70.1895}. Depending on the properties of the Hilbert space of the involved systems, the teleportation can be classified into discrete- and continuous-variable \cite{PhysRevLett.80.869,doi:10.1126/science.282.5389.706,PhysRevLett.115.180502,PhysRevLett.123.070505} quantum teleportation.

In the case of discrete-variable teleportation, both the quantum channel and the unknown state are constructed of two-level systems (TLSs) \cite{ma2012quantum} [see Fig. \ref{flow diagram}(a)]. Consider that Alice wants to transfer an unknown state $\ket{\varphi_1}=a\ket{e_1}+b\ket{g_1}$ carried by the first TLS on her hands to Bob, where $|e_l\rangle$ and $|g_l\rangle$ are the excited and ground states of the $l$th TLS, respectively. She first shares an entangled pair of TLSs in the state $\ket{\Phi_{23}}=\frac{1}{\sqrt{2}}(\ket{e_2e_3}+\ket{g_2g_3})$ as a quantum channel with Bob. The quantum channel evolves with time under a unitary dynamics governed by the Hamiltonian $\hat{H}_0=\omega_0\sum_{l=2}^3\hat{\sigma}_l^\dag\hat{\sigma}_l$, with $\hat{\sigma}_l=|g_l\rangle\langle e_l|$. Thus, the state of the tripartite TLS dynamically evolves to
\begin{eqnarray}
\ket{\Phi_\text{T}(t)}&=&2^{-1}[|\Phi^+_{12}\rangle\hat{I}_3+|\Phi^-_{12}\rangle\hat{\sigma}^z_{3}+|\Psi^+_{12}\rangle\hat{\sigma}_3^x\nonumber\\
 &&+|\Psi^-_{12}\rangle i\hat{\sigma}_3^y](ae^{-i2\omega_0t}|e_3\rangle+|g_3\rangle),
\end{eqnarray}
where $|\Phi^\pm\rangle=(|ee\rangle\pm|gg\rangle)/\sqrt{2}$ and $|\Psi^\pm\rangle=(|eg\rangle\pm|ge\rangle)/\sqrt{2}$ are the four Bell states \cite{PhysRevA.59.3295}. Then, Alice makes a Bell-state measurement to her two TLSs. If the $k$th Bell state $|\text{Bell}_{12}^{(k)}\rangle$, with $|\text{Bell}\rangle\in\{|\Phi^\pm\rangle,|\Psi^\pm\rangle\}$, is obtained, the state of Bob's TLS collapses to $\langle\text{Bell}_{12}^{(k)}|\Phi_{123}(t)\rangle$. Alice tells Bob  her measurement result $k$ via the classical channel. Finally, Bob performs a local operation $\hat{U}_3^{(k)}$ on his TLS according to the result $k$. If the measurement result is $|\Phi_{12}^+\rangle$, $|\Phi_{12}^-\rangle$, $|\Psi_{12}^+\rangle$, and $|\Psi_{12}^-\rangle$, he chooses the operation $\hat{I}$, $\hat{\sigma}^z$, $\hat{\sigma}^x$, and $-i\hat{\sigma}^y$, respectively. Bob's final state then becomes
\begin{equation}\label{3}
|\varphi_{3}^{(k)}\rangle=P_k^{-1/2}\hat{U}_3^{(k)}\langle \text{Bell}^{(k)}_{12}|\Phi_\text{T}(t)\rangle,
\end{equation}
where $P_k=|\langle \text{Bell}_{12}^{(k)}|\Phi_\text{T}(t)\rangle|^2$. Setting $a= \cos(\theta/2)$ and $b = \sin(\theta/2)\mathit{e}^{\mathit{i}\phi}$, we can evaluate the average fidelity as
\begin{eqnarray}
\bar{F}&=&\int_0^{\pi}\frac{\sin\theta}{4\pi} d\theta\int_0^{2\pi}d\phi\sum_{k=1}^4P_k|\langle\varphi|\varphi_{3}^{(k)}\rangle|^2\nonumber\\&=&[2+\cos(2\omega_0t)]/3.\label{4}
\end{eqnarray}
Thus, a maximal average fidelity $\max_t\bar{F}(t)=1$ is achieved at times $t=n\pi/\omega_0$, with $n$ being positive integers.

In the case of continuous-variable teleportation, the information is encoded in quantized optical fields [see Fig. \ref{flow diagram}(b)]. When Alice wants to transfer an unknown coherent state $|\varphi_1\rangle=\hat{D}_1(\alpha)|0_1\rangle$, with $\hat{D}_1(\alpha)=e^{\alpha\hat{a}_1^\dag-\alpha^*\hat{a}_1^\dag}$, to Bob, she first shares a two-mode squeezed vacuum state $|\Phi_{23}\rangle=\exp [r(\hat{a}_2\hat{a}_3-\hat{a}^\dag_2\hat{a}_3^\dag)]|0_20_3\rangle$ as a quantum channel with Bob \cite{RevModPhys.84.621}. Considering the unitary dynamics governed by $\hat{H}_0=\omega_0\sum_{l=2}^3\hat{a}_l^\dag\hat{a}_l$ of the quantum channel, the state of the total system reads
\begin{eqnarray}
|\Phi_\text{T}(t)\rangle=\cosh^{-1} re^{\frac{-|\alpha|^2}{2}-\alpha\hat{a}_1^\dag-\tanh re^{-2i\omega_0 t}\hat{a}_2^\dag\hat{a}_3^\dag}|0_10_20_3\rangle.
\end{eqnarray}
Then, Alice makes a homodyne measurement to the two optical fields on her hands. It consists of the following two steps. The first step Alice needs to do is to couple the optical fields $\hat{a}_1$ and $\hat{a}_2$ by a 50:50 beam splitter, which can be described by $\hat{V}=\exp[\frac{\pi}{4}(\hat{a}_1^\dag\hat{a}_2-\hat{a}_2^\dag\hat{a}_1)]$. The second one is to measure two commuting quadrature operators $\hat{X}_1=(\hat{a}_1+\hat{a}_1^\dag)/ 2$ and $\hat{P}_2=(\hat{a}_2-\hat{a}_2^\dag)/(2i)$. The homodyne measurement, with the results $x_1$ and $p_2$, collapses the state of Bob's optical field into
\begin{eqnarray}
{\langle x_1p_2|\hat{V} |\Phi_\text{T}(t)\rangle\over\sqrt{P_z}}&=&c(z)e^{(\alpha-\sqrt{2}z)e^{-i2\omega_0t}\tanh r\hat{a}^\dag_3}|0_3\rangle,~\label{fdfd}
\end{eqnarray}
where $z=x_1-ip_2$, $c(z)={e^{-{|\alpha|^2\over2}-|z|^2+\sqrt{2}\alpha z^*}\over \sqrt{2^{-1}\pi P_z}\cosh r}$, $\hat{X}_1|x_1\rangle=x_1|x_1\rangle$, $\hat{P}_2|p_2\rangle=p_2|p_2\rangle$, and $P_z=|\langle x_1p_2|\hat{V} |\Phi_\text{T}(t)\rangle|^2$. In the derivation of Eq. \eqref{fdfd}, $\langle p_1x_2|0_10_2\rangle=({2\over\pi})^{1/2}e^{-x^2-p^2}$ has been used. Alice tells Bob her measurement result $z$ via the classical channel. Finally, Bob makes a displacement operation $\hat{D}_3(\sqrt{2}z)$ on his optical field to convert Eq. \eqref{fdfd} into
\begin{eqnarray}
|\varphi^{(z)}_3\rangle=c(z)\big|[\sqrt{2}z+(\alpha-\sqrt{2}z)e^{-i2\omega_0t}\tanh r]_3\big\rangle.
\end{eqnarray} If a series of teleported states are in sequence given to Alice and the homodyne detector is able to respond to all of eigenvalues of the quadrature operators in the teleportation, then the average fidelity over all the measurement results is \cite{LiFu-Li:14}
\begin{eqnarray}
\bar{F}(t)=\int dx_1 dp_2 P_z|\langle \alpha|\varphi^{(z)}_3\rangle|^2={1/2\cosh^{-2}r\over 1-\tanh r\cos(2\omega_0 t)}.\label{cvqtf}
\end{eqnarray}
A maximal average fidelity of $\max_t\bar{F}(t)=(1+e^{-2r})^{-1}$ is achieved when $t=n\pi/\omega_0$. It tends to 1 when $r$ approaches infinity.

\section{Noisy effects}\label{noiseff}
In practice, the quantum channel is inevitably influenced by the noise-induced decoherence. It causes the degradation of the entanglement and deteriorates the performance of quantum teleportation. Depending on whether the system in the channel has energy exchange with the noise or not, the decoherence is classified into dissipation and dephasing. The description of decoherence is based on the idea of the open system. Starting from the unitary dynamics of the total system formed by the system and its noise and tracing over the degrees of freedom the noise, the dynamics of the open system is achieved. A widely used approximation during this procedure is the Born-Markovian approximation \cite{book_open}. The spontaneous emission of the two two-level systems and the photon loss involved in the quantum channel are two main decoherence sources in quantum teleportation. Conventionally, they were phenomenologically described by the Kraus-operator representation for the discrete-variable channel \cite{PhysRevA.90.042332,PhysRevA.92.012338} and by introducing an imperfect beam splitter for the continuous-variable channel \cite{PhysRevA.65.022310,PhysRevA.105.062407,PhysRevLett.108.130402}, both of which are equivalent to the Born-Markovian approximate description \cite{yeo2009effects,PhysRevLett.108.130402}. We here will go beyond this approximation \cite{Rivas_2014,RevModPhys.88.021002,LI20181} and exactly evaluate the effects of local dissipative noises on the two types of schemes of quantum teleportation. In Ref. \cite{PhysRevA.108.062406}, the non-Markovian effect is simulated by coupling the quantum system to an ancillary system, which feels a white-noise reservoir. Different from this, we  microscopically investigate the non-Markovian effect caused by the direct coupling of the quantum channel to the reservoir.

The dissipative noise can be described by a bosonic reservoir. Consider that each subsystem of the quantum channel feels a local reservoir. The Hamiltonian for both the discrete- and continuous-variable quantum channels can be universally written as $\hat{H}=\sum_{l=2}^3\hat{H}_l$, with
\begin{equation}
\hat{H}_l=\omega_0\hat{o}_l^\dag\hat{o}_l+\sum_{k}[\omega_{k}\hat{b}_{l,k}^{\dagger}\hat{b}_{l,k}+ (g_k\hat{b}_{l,k}^{\dagger}\hat{o}_l+ \text{h.c.})],\label{eq:Hamiltonian}
\end{equation}
where $\hat{b}_{l,k}$ is the annihilation operator of the $k$th mode with frequency $\omega_k$ of the reservoir felt by the $l$th subsystem of the quantum channel, $g_k$ is the coupling strength, $\hat{o}=\hat{\sigma}$ for the discrete-variable teleportation, and $\hat{o}=\hat{a}$ for the continuous-variable teleportation. Commonly, the coupling strength is further characterized by the  spectral density $J(\omega)=\sum_k\left|g_k\right|^2\delta(\omega-\omega_k)$. It may have the form $J(\omega)=\eta\omega^s\omega_c^{1-s}e^{-\omega/\omega_c}$, where $\eta$ is a dimensionless coupling constant, $\omega_c$ is a cut-off frequency to avoid infrared catastrophe, and $s$ is the Ohmicity parameter. The reservoir is classified into sub-Ohmic for $0< s <1$, Ohmic for $s = 1$, and super-Ohmic for $s>1$ \cite{leggett1987dynamics}. Under the condition that the reservoirs are initially in the vacuum state, one can trace out the degrees of freedom of the reservoirs exactly. The dynamics governed by Eq. \eqref{eq:Hamiltonian} for the discrete-variable system is exactly solvable because only the Hilbert subspaces with the total excitation numbers $N=0$ and $1$ are involved \cite{PhysRevA.81.052330}. The exact dynamics for the continuous-variable system is obtainable by the Feynman-Vernon's influence-functional theory in the coherent-state representation \cite{an2007non1,An_2009}. We obtain the non-Markovian master equation as
\begin{eqnarray} \label{9}
\dot{\rho}(t) &=&\sum_{l=2,3}\{-i\Omega(t)[\hat{o}^\dag_{l}\hat{o}_l,\rho(t)] + \Gamma(t)\check{\mathcal{L}}_l\rho(t)] \},
\end{eqnarray}
where $\rho(t)$ is the reduced density matrix of the quantum channel formed by the subsystems $\hat{o}_2$ and $\hat{o}_3$, $\check{\mathcal{L}}_l\rho(t)=2\hat{o}_l\rho(t)\hat{o}^\dag_l-[\hat{o}^\dag_l\hat{o}_l,\rho(t)]_+$ is the Lindblad superoperator, $\Gamma(t)\equiv-\text{Re}[\dot{u}(t)/u(t)]$ is the decay rate, and $\Omega(t)\equiv-\text{Im}[\dot{u}(t)/u(t)]$ is the renormalized frequency. The time-dependent function $u(t)$ is determined by
\begin{equation}\label{10}
\dot{u}(t) + \mathit{i}\omega_0u(t) + \int_0^t d\tau\mu(t-\tau)u(\tau) = 0,
\end{equation}
where $\mu(x) \equiv \int_0^\infty d\omega J(\omega)e^{-\mathit{i}\omega x}$ is the noise correlation function and the initial condition is $u(0)=1$.

In the special case when the system-reservoir coupling is weak and the characteristic time scale of the correlation function of the reservoirs is much smaller than that of the system, we can apply the Born-Markovian approximation to Eq. \eqref{10} by replacing $u(\tau)$ by $u(t)$ and extending the upper limit of the integration from $t$ to infinity. Then using the identity $\lim_{t\rightarrow\infty}\int_0^t d\tau e^{-i(\omega-\omega_0)\tau}=\pi\delta(\omega-\omega_0)+i{\mathcal{P}\over\omega_0-\omega}$, with $\mathcal{P}$ being the Cauchy principal value, the Born-Markovian approximate solution of Eq. \eqref{10} reads
\begin{equation}\label{13}
u_\text{BMA}(t) \backsimeq e^{-[\kappa+i(\omega_0+\Delta_{\omega_0})]t},\end{equation}
where $\kappa =\pi J(\omega_0)$ and $\Delta_{\omega_0} = \mathcal{P}\int_0^{\infty}\frac{J(\omega)}{\omega_0 - \omega}d\omega$.

In the general non-Markovian case, the exact evaluation of the average fidelity $\bar{F}(t)$ needs the numerical solving of Eq. \eqref{10}. However, via analyzing the long-time behavior of $u(t)$, we can obtain an analytical form of $\bar{F}(t)$ in the steady-state limit of the quantum channel. This is helpful for us to build up a clear physical picture on the performance of our quantum-teleportation schemes under the impact of the local reservoirs. A Laplace transform to Eq. \eqref{10} results in $\tilde{u}(p) = [p+i\omega_0+\int_0^\infty\frac{J(\omega)}{p+i\omega}d\omega]^{-1}$. $u(t)$ is obtained by the inverse Laplace transform to $\tilde{u}(p)$, which can be done by finding the poles from
\begin{equation}\label{E}
Y(E) \equiv \omega_0 - \int_0^{\infty}\frac{J(\omega)}{\omega - E}d\omega = E, ~(E=ip).
\end{equation}
It is interesting to find that the roots of Eq. \eqref{E} are just the eigenenergies of each subsystem and its local reservoir in the single-excitation subspace. To be specific, we expand the eigenstate of $\hat{H}_l$ as $\ket{\Phi} = (x\hat{o}_l^{\dagger}+\sum_k y_k\hat{b}_k^{\dagger})|\varnothing_l,\{0_k\}\rangle$, where $|\varnothing\rangle=|g\rangle$ and $|0\rangle$ for the discrete- and continuous-variable ones, respectively. Substituting it into $\hat{H}_l\ket{\Phi} = E\ket{\Phi}$ with $E$ being the eigenenergy, we can readily find that the roots $E$ satisfy the same equation as Eq. \eqref{E}. This result implies that the decoherence dynamics of both of the quantum channels is essentially determined by the energy-spectrum characteristic of each local subsystem-reservoir system. It gives us an insightful message to control the decoherence dynamics of the quantum channels by artificially engineering the structure of the energy spectrum. Because $Y(E)$ is a monotonically decreasing function in the regime $E < 0$, Eq. \eqref{E} has one and only one isolated root $E_b$ provided $Y(0)<0$. We call the eigenstate of $E_b$ the bound state. On the other hand, since $Y(E)$ is not well analytic in the regime $E>0$ due to the divergence of its integration, Eq. \eqref{E} has infinite roots in this regime. They form a continuous energy band. Then, after applying the inverse Laplace transform using the contour integration and  the residue theorem, we obtain \cite{wu2021non}
\begin{equation}\label{16}
u(t) = Ze^{-iE_b t} + \int_0^{\infty}\frac{J(E)e^{-iEt}dE}{(E-\omega_0-\Delta_E)^2+[\pi J(E)]^2},
\end{equation}
where the first term with $Z \equiv [1+\int_0^{\infty}\frac{J(\omega)}{(E_b-\omega)^2}d\omega]^{-1}$ is from the potentially formed bound state and the second term is from the band energies. Contributed by the branch cut of the contour, the second term approaches zero in the long-time regime due to the out-of-phase interference of the continuously changing oscillation frequency $E$. Thus, if the bound state is absent, we have $u(\infty)=0$ characterizing a complete decoherence; while if the bound state energy is formed, we have $u(\infty)\simeq Ze^{-iE_b t}$ implying a decoherence suppression. The condition $Y(0)<0$, under which the bound state is formed, can be evaluated for the Ohmic-family spectral density  as $\omega_0 - 2\eta\omega_c\gamma(s) \leq 0$, where $\gamma(s)$ is the Euler's gamma function. Thus, via efficiently engineering the system frequency $\omega_0$ and the parameters in the spectral density to form the bound state, we can suppress the decoherence in quantum teleportation. To verify this expectation, we investigate the decoherence effects on the two types of quantum teleportation schemes in the following.

\subsection{Discrete-variable case}
Solving the master equation \eqref{9} with $\hat{o}=\hat{\sigma}$ under the initial state $\rho_{23}(0) = \ket{\Phi_{23}}\bra{\Phi_{23}}$, we obtain
\begin{eqnarray}\label{11}
\rho_{23}(t) &=&\{[P_t|e\rangle\langle e|+(1-P_t)|g\rangle\langle g|]^{\otimes2} + |g\rangle\langle g|^{\otimes2} \notag \\
 &&+[u^2(t)|e\rangle\langle g|^{\otimes2}+\text{h.c.}]\}/2 ,
\end{eqnarray}where $P_t=|u(t)|^2$. Repeating the similar procedure as the ideal case, we get the average fidelity as
\begin{eqnarray}
\bar{F}(t)&=&\int_0^{\pi}{\sin\theta \over 4\pi}d\theta\int_0^{2\pi}d\phi\sum_{k=1}^4\langle\varphi|\hat{U}_3^{(k)}\langle \text{Bell}^{(k)}_{12}|\varphi_1\rangle\langle \varphi_1|\notag\\
&&\otimes\rho_{23}(t)|\text{Bell}^{(k)}_{12}\rangle\hat{U}_3^{(k)\dagger}|\varphi\rangle\nonumber\\
&=&\{2+|u(t)|^2(|u(t)|^2-1)+\text{Re}[u(t)^2]\}/3.\label{12}
\end{eqnarray}
It is readily checked that $u(t)=e^{-i\omega_0t}$ and thus $\bar{F}(t)$ tends to the ideal result of Eq. \eqref{4} in the noiseless limit. Substituting the Born-Markovian approximation result \eqref{13} into Eq. \eqref{12}, we also obtain
\begin{equation}\label{14}
\bar{F}_\text{BMA}(t)= \{ 2+e^{-2\kappa  t}[e^{-2\kappa t}-2\sin^2(\omega_0t)]\}/3,
\end{equation}where the frequency shift $\Delta_{\omega_0}$ has been renormalized into the bare frequency $\omega_0$. $ \bar{F}_\text{BMA}(t)$ tends to $2/3$ with time, which is just the classical-communication limit \cite{PhysRevLett.74.1259}. It means that the quantum advantage of teleportation is completely destroyed by the Markovian dissipative noises. A similar result was also reported in many previous works.

In the non-Markovian case, it is natural to expect that $\bar{F}(t)$ tends to $2/3$ with time too when no bound state is formed in the energy spectrum of each TLS and its local reservoir. We focus on the situation in the presence of the bound state. Substituting $\lim_{t\rightarrow \infty}u(t)=Ze^{-iE_bt}$ in the situation into Eq. \eqref{12}, we obtain
\begin{equation}\label{17}
\lim_{t\rightarrow\infty}\bar{F}(t) =\{2+Z^2[Z^2-2\sin^2(E_bt)]\}/3.
\end{equation}
It is remarkable to find that $\max\lim_{t\rightarrow\infty}\bar{F}(t)=(2+Z^4)/3$ is always larger than the classical fidelity $2/3$. Therefore, the quantum superiority of the teleportation is retrieved in the non-Markvoain dynamics as long as the bound state is formed.

\subsection{Continuous-variable case}
Within the framework of the path-integral influence-function method, the solution of the master equation \eqref{9} for the continuous-variable quantum channel is  \cite{an2007non1,An_2009}
\begin{eqnarray}
\rho (\pmb{\bar{\alpha}}_{f},\pmb{\alpha }_{f}^{\prime };t)&=&\int d\mu (\pmb{\alpha }_{i})d\mu (\pmb{\alpha }_{i}^{\prime })\mathcal{J}(\pmb{\bar{\alpha}}_{f},\pmb{\alpha }_{f}^{\prime};t|\pmb{\alpha}_{i},\bar{\pmb{\alpha }}_{i}^{\prime };0)  \notag  \label{rout} \\
&&\times \rho (\pmb{\bar{\alpha}}_{i},\pmb{\alpha }_{i}^{\prime };0),\label{tmdps}
\end{eqnarray}
where $\rho (\pmb{\bar{\alpha}}_{f},\pmb{\alpha }_{f}^{\prime};t)=\langle \pmb{\alpha }_{f}|\rho_{23} (t)|\pmb{\alpha }_{f}^{\prime }\rangle $ is the reduced density matrix expressed in coherent-state representation and $\mathcal{J}(\pmb{\bar{\alpha}}_{f},\pmb{\alpha }_{f}^{\prime };t|\pmb{\bar{\alpha}}_{i},\pmb{\alpha }_{i}^{\prime };0)$ is the propagating function. We have used the coherent-state representation $|\pmb{\alpha }\rangle =\prod_{l=2}^{3}|\alpha _{l}\rangle$, with $|\alpha_{l}\rangle =\exp (\alpha _{l}\hat{a}_{l}^{\dagger })|0_{l}\rangle$, which obeys the resolution of identity $\int d\mu \left( \pmb{\alpha }\right) |\pmb{\alpha }\rangle \langle \pmb{\alpha }|=1$ with the integration measures $d\mu \left( \pmb{\alpha }\right) \equiv\prod_{l}e^{-\bar{\alpha}_l\alpha_l}\frac{d^{2}\alpha _{l}}{\pi }$. $\bar{\pmb{\alpha }}$ denotes the complex conjugate of $\pmb{\alpha }$. The propagating function reads
\begin{eqnarray}
&&\mathcal{J}(\pmb{\bar{\alpha}}_{f},\pmb{\alpha }_{f}^{\prime};t|\pmb{\alpha}_{i},\bar{\pmb{\alpha }}_{i}^{\prime };0)=\exp\big\{\sum_{l=2,3}\big[u(t)\bar{\alpha}_{lf}\alpha _{li}  \notag \\
&&~~~~~+\bar{u}(t)\bar{\alpha}_{li}^{\prime }\alpha _{lf}^{\prime}+[1-\left\vert u(t)\right\vert ^{2}]\bar{\alpha}_{li}^{\prime }\alpha_{li}\big]\big\}.  \label{prord}
\end{eqnarray}The coherent-state representation of the two-mode squeezed state as the initial state is
\begin{eqnarray}\label{iniat}
\rho (\boldsymbol{\bar{\alpha}}_i,\boldsymbol{\alpha }^{\prime }_i;0)=\cosh ^{-2}r\exp [-\tanh  r (\bar{\alpha}_{1}\bar{\alpha}_{2}+\alpha _{1}^{\prime }\alpha _{2}^{\prime })].
\end{eqnarray}
Substituting Eqs. \eqref{iniat} and \eqref{prord} into Eq. \eqref{tmdps} and performing the Gaussian integration, we obtain
\begin{equation}
\rho (\pmb{\bar{\alpha}}_{f},\pmb{\alpha }_{f}^{\prime };t)=a\exp\normalsize[\sum_{l\neq l^{\prime }}(\frac{b}{2}\bar{\alpha}_{lf}\bar{\alpha}_{l^{\prime }f}+c\bar{\alpha}_{lf}\alpha _{lf}^{\prime }+\frac{b^{\ast }}{2}\alpha _{lf}^{\prime }\alpha _{l^{\prime }f}^{\prime })],
\end{equation}
with $a=x\cosh^{-2}r$, $b=-xu(t)^2\tanh r$, $c=x|u(t)|^2(1-|u(t)|^2)\tanh^2 r$, and $x=[1-(1-|u(t)|^2)^2\tanh^2r]^{-1}$. A procedure similar to that of the ideal case leads to the average fidelity as
\begin{eqnarray}
\bar{F}(t)&=&\int dx_1dp_2\langle\varphi|\hat{D}_3(\sqrt{2}z)\langle x_1p_2|\hat{V} |\varphi_1\rangle\langle\varphi_1|\rho_{23}(t)\hat{V}^\dag\nonumber\\
&&\otimes  |x_1p_2\rangle  \hat{D}^\dag_3(\sqrt{2}z)|\varphi\rangle={a\over 2}[1+\text{Re}(b)-c]^{-1}.\label{extfdc}
\end{eqnarray}
In the noiseless limit, $u(t)=e^{-i\omega_0t}$ and thus $a=\cosh^{-2}r$, $b=-e^{-2i\omega_0t}\tanh r$, and $c=0$, which readily reduces $\bar{F}(t)$ to Eq. \eqref{cvqtf}. Considering the Born-Markovian approximation solution \eqref{13}, we obtain
\begin{equation}\label{mdss}
\bar{F}_\text{BMA}(t)=\{2+\sinh(2r)e^{-2\kappa t}[\tanh r-\cos(2\omega_0t)]\}^{-1}.
\end{equation}
Equation \eqref{mdss} tends to $1/2$ in the long-time condition, which is the classical (i.e., no entanglement) bound for teleporting coherent states \cite{PhysRevA.64.022321,PhysRevLett.94.150503}. Therefore, the quantum advantage of teleportation again is destroyed by the Markovian noises.

\begin{figure}[tbp]
\centering
\includegraphics[width=\columnwidth]{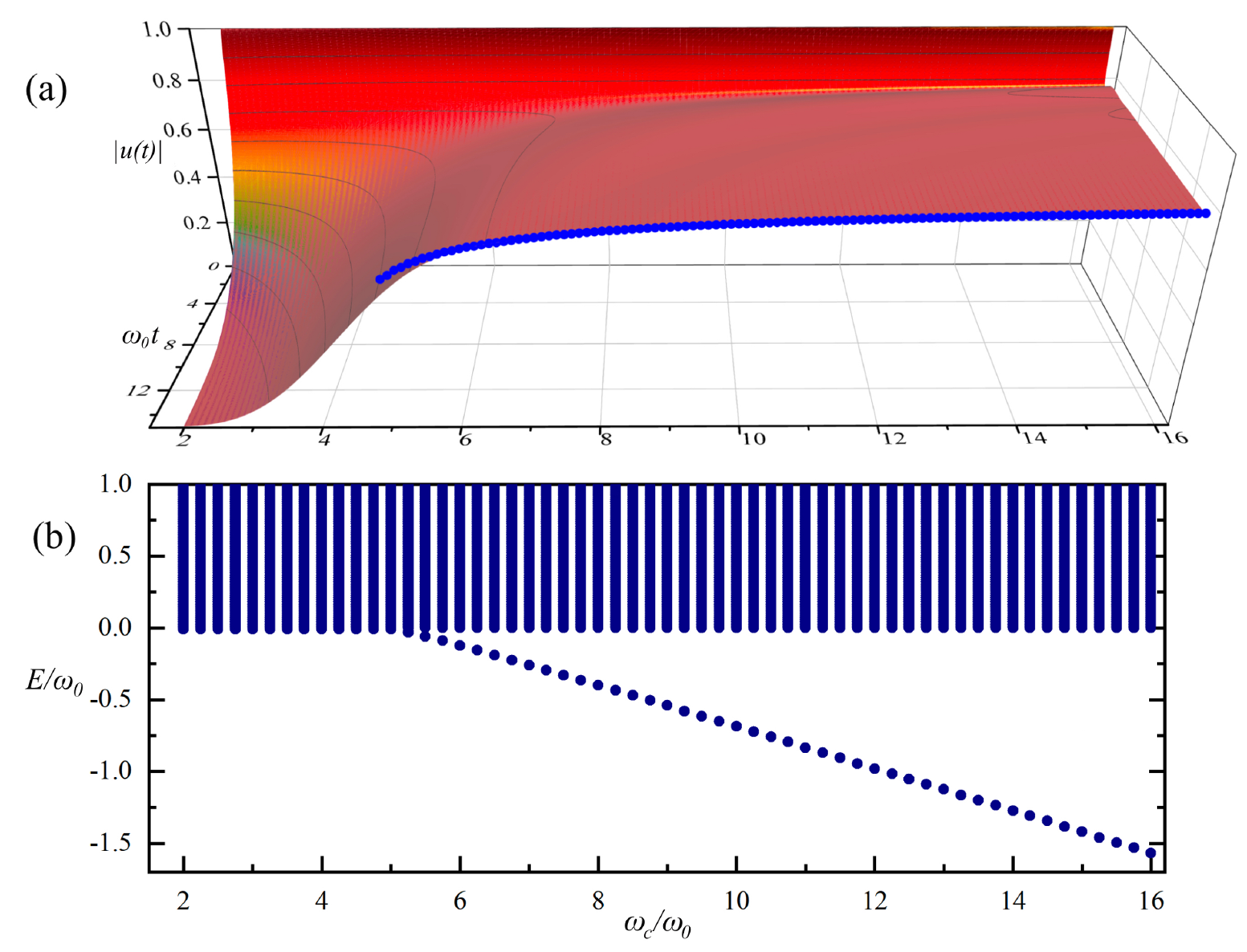}
\caption{(a) Evolution of $|u(t)|$ obtained via numerically solving Eq. \eqref{10} and $Z$ (blue line) in different $\omega_c$. (b) Energy spectrum of the total system formed by the system and the reservoir in different $\omega_c$ obtained via numerically solving Eq. \eqref{E}. We use $\eta=0.2$ and $s=1$, where the bound state is formed when $\omega_c>5\omega_0$.  }\label{energspectrm}
\end{figure}
In the general non-Markovian dynamics, substituting $\lim_{t\rightarrow \infty}u(t)=Ze^{-iE_bt}$ in the presence of the bound state into Eq. \eqref{extfdc}, we have
\begin{equation}\label{32}
\lim_{t\rightarrow\infty}\bar{F}(t)=\{2+\sinh(2r)Z^2[\tanh r-\cos(2E_bt)]\}^{-1}.
\end{equation}
Its maximum $\max \lim_{t\rightarrow\infty}\bar{F}(t)=[2-Z^2(1-e^{-2r})]^{-1}$ is achieved when $t=n\pi/(2E_b)$ with $n$ being odd numbers. It is always greater than the Born-Markovian approximate result, i.e., $1/2$. Therefore, being similar to the discrete-variable case, we can retrieve the quantum superiority of the noisy teleportation in the non-Markovian dynamics by engineering the formation of the bound state. It is noted that our result is different from the one in Ref. \cite{PhysRevA.108.062406}, where  the non-Markovian effect just slows down the deterioration of the performance of the quantum teleportation and the average fidelity still becomes smaller than $1/2$ in the long-time condition. In sharp contrast to this, our result reveals a mechanism in the non-Markovian dynamics to keep the quantum superiority of the average fidelity till the long-time steady state. The validity of this mechanism is guaranteed by the feature of the energy spectrum of the total system formed by the quantum channel and its reservoir, i.e., the formation of the bound state.

\begin{figure}[tbp]
\centering
\includegraphics[width=\columnwidth]{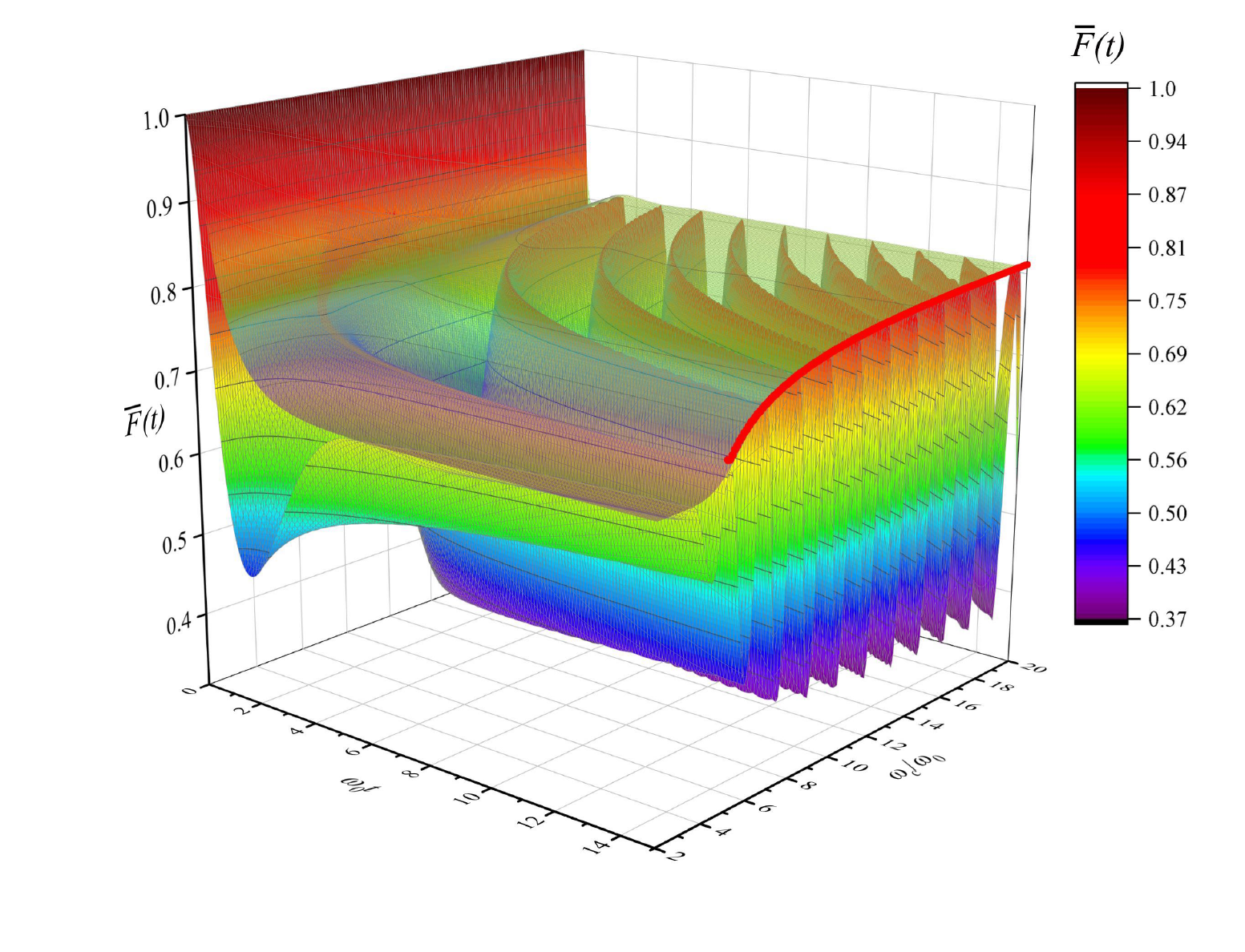}
\caption{Evolution of the average fidelity $\bar{F}(t)$ of Eq. \eqref{12} (the lower surface), its transient maxima in Eq. \eqref{17} (the upper surface), and its steady-state maxima $\max\lim_{t\rightarrow\infty}\bar{F}(t)=(2+Z^4)/3$ (red line) in different $\omega_c$ in the discrete-variable teleportation case. The parameters are the same as those in Fig. \ref{energspectrm}.}\label{dscrt}
\end{figure}

\section{Numerical results}\label{numres}

To verify our noise mitigation mechanism in quantum teleportation, we first plot in Fig. \ref{energspectrm}(a) the non-Markovian evolution of $|u(t)|$ in different $\omega_c$ by choosing $s=1$ and $\eta=0.2$. Two typical behaviors are observed. For $\omega_c<5\omega_0$, $|u(t)|$ decays exclusively to 0. When $\omega_c>5\omega_0$, $|u(t)|$ halts to decay and tends to a finite value. It is interesting to see that this finite value $|u(\infty)|$ matches exactly with $Z$ evaluated from the bound state [see the blue line in Fig. \ref{energspectrm}(a)]. Completely different from the Markovian result, such decoherence suppression is a distinctive feature of our non-Markovian dynamics. The energy spectrum of the total system formed by the each subsystem of the quantum channel and its local reservoir in Fig. \ref{energspectrm}(b) reveals that a bound state is formed when $\omega_c>5\omega_0$, which coincided exactly with the regime where $|u(\infty)|$ tends to a finite value in Fig. \ref{energspectrm}(a). The result demonstrates the decisive role of the bound state in suppressing the non-Markovian decoherence of the quantum channel. This gives us an insightful inspiration to retrieve the superiority of quantum teleportation in the presence of the practical noise by engineering the formation of the bound state.

\begin{figure}[tbp]
\centering
\includegraphics[width=\columnwidth]{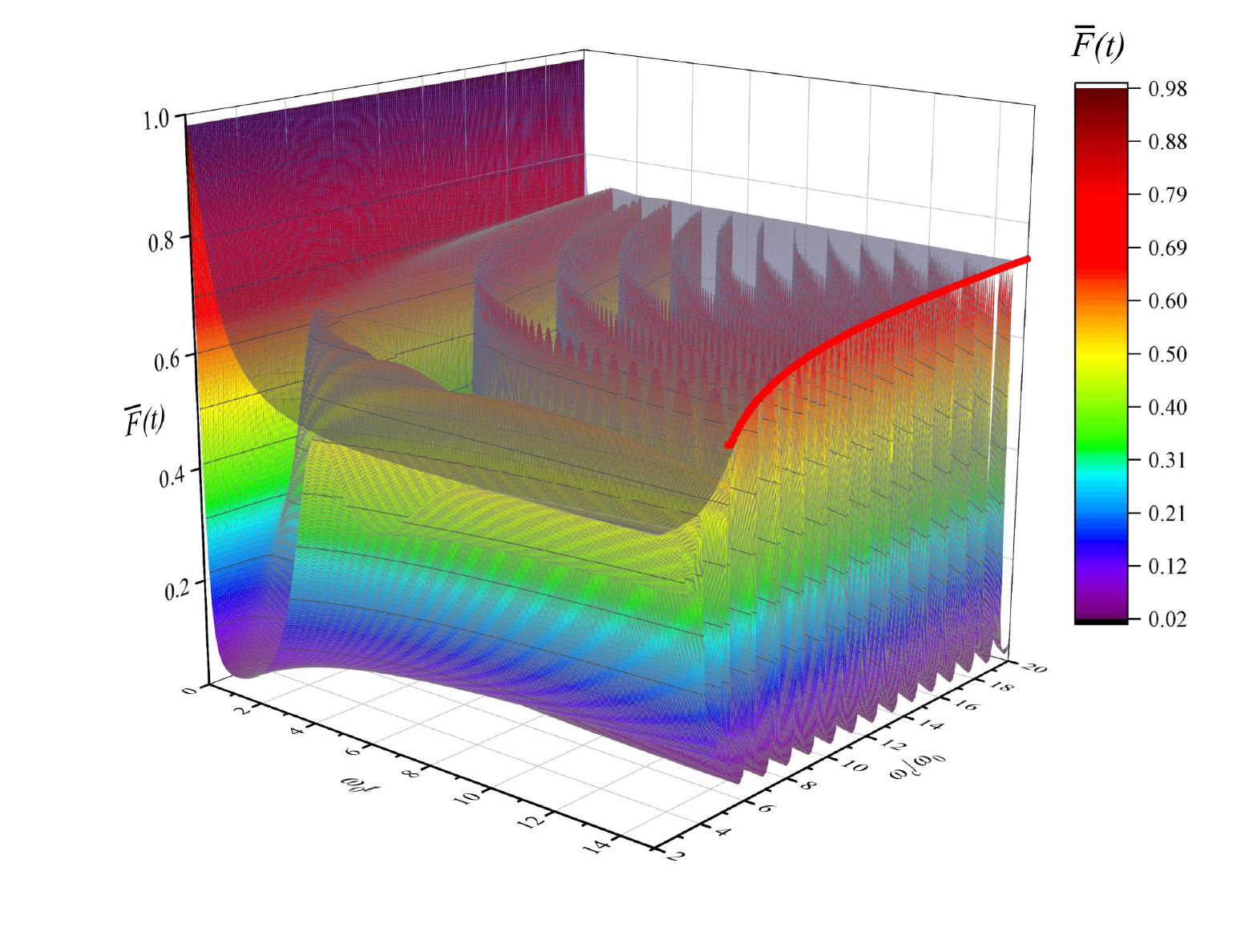}
\caption{Evolution of the average fidelity $\bar{F}(t)$ of Eq. \eqref{extfdc} (the lower surface), its transient maxima in Eq. \eqref{32} (the upper surface), and its steady-state maxima $\max \lim_{t\rightarrow\infty}\bar{F}(t)=[2-Z^2(1-e^{-2r})]^{-1}$ (red line) in different $\omega_c$ in the continuous-variable teleportation case. $r=2$ and other parameters are the same as those in Fig. \ref{energspectrm}.}\label{ctnvar}
\end{figure}

Figure \ref{dscrt} shows the evolution of the average fidelity $\bar{F}(t)$ in Eq. \eqref{12} of the discrete-variable quantum teleportation. It clearly indicates that $\bar{F}(t)$ tends to $2/3$ in the absence of the bound state when $\omega_c<5\omega_0$. As long as the bound state is present, the transient maxima of $\bar{F}$, which matches with the analytical result in Eq. \eqref{17} evaluated from the bound state, approaches $\max\lim_{t\rightarrow\infty}\bar{F}(t)=(2+Z^4)/3$ in the long-time limit (see the red line in Fig. \ref{dscrt}). It fully reveals the recovery of the quantum advantage of the teleportation due to the formation of the bound state in the non-Markovian decoherence dynamics. The result in the continuous-variable case in Fig. \ref{ctnvar} also confirms our expectation. It shows that when $\omega_c<5 \omega_0$, $\bar{F}(t)$ approaches 1/2, which shows no difference from the Markovian approximate result in Eq. \eqref{mdss}. When $\omega_c>5\omega_0$, the transient maxima of $\bar{F}(t)$ tend to the values evaluated from the analytical Eq. \eqref{32}. Its long-time maxima fit very well with $\max \lim_{t\rightarrow\infty}\bar{F}(t)=[2-Z^2(1-e^{-2r})]^{-1}$, which surpasses 1/2 and partially retrieves the quantum advantage from the destruction of noise.

These results demonstrate the important role of the bound state and non-Markovian dynamics in recovering the quantum advantage of the discrete- and continuous-variable teleportation in the noisy situation. Such an amazing result is caused by the suppressed decoherence induced by the bound state in the non-Markovian dynamics. Therefore, we can mitigate the loss of the quantum advantage of teleportation \cite{lee2000entanglement,oh2002fidelity,jung2008greenberger,rao2008teleportation,yeo2009effects} via engineering the formation of the bound state. It should be noted that the distinguished roles played by the bound state in other quantum protocols have been reported \cite{PhysRevLett.123.040402,PhysRevLett.131.050801,PhysRevLett.132.090401}.

\begin{figure}[tbp]
\centering
\includegraphics[width=\columnwidth]{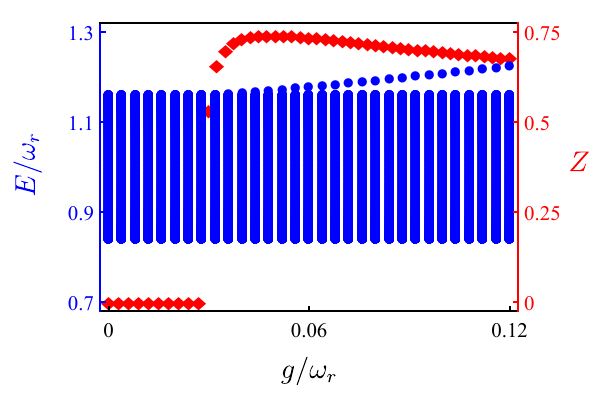}
\caption{Energy spectrum of a superconductor qubit or an LC resonator interacting with a coupled resonator array and $Z$ in different $g$ obtained by numerically solving Eq. \eqref{E}. We use $N=500$, $\omega_0=1.15\omega_r$, and $\xi=0.08\omega_r$.  }\label{carrf}
\end{figure}
\section{Physical realization}\label{phdrd}
It should be emphasized that the bound-state-favored superiority in our quantum teleportation of both discrete- and continuous-variable systems is independent of the explicit form of the spectral density. Although only the Ohmic-family spectral density is displayed, our result can be generalized to other cases without difficulty. The most proper system to verify our result is structured reservoirs, whose non-Markovian effect is generally strong. Inspired by the experimental realization of quantum teleportation in the circuit quantum electrodynamics setup for both the discrete-variable \cite{PhysRevLett.108.040502} and the continuous-variable \cite{doi:10.1126/sciadv.abk0891} systems, we propose the following platform to verify our results. The discrete-variable quantum channel is formed by two superconductor qubits. The continuous-variable quantum channel is formed by two quantized microwave fields in two LC resonators. Each subsystem of the two quantum channels interacts with a coupled resonator array as the reservoir. The Hamiltonians are
\begin{eqnarray}
\hat{H}_{l,\text{R}}&=&\sum_{N=1}^N\omega_r\hat{\tilde b}_{l,j}^\dag\hat{\tilde b}_{l,j}+\sum_{N=1}^{N-1}\xi(\hat{\tilde b}_{l,j}^\dag\hat{\tilde b}_{l,j+1}+\text{H.c.}),\label{hrcc} \\
\hat{H}_{l,\text{int}}&=&g(\hat{o}_l^\dag\hat{\tilde b}_{l,1}+\text{H.c.}),
\end{eqnarray}where $\omega_r$ is the frequency of the resonator, $\xi$ is the coupling strength between the nearest-neighbor resonators in the array, and $g$ is the coupling strength between the $l$th subsystem and its own coupled resonator array. Equation \eqref{hrcc} is rewritten in the momentum space as $\hat{H}_{l,\text{R}}=\sum_k\omega_k\hat{b}_{l,k}^\dag\hat{b}_{l,k}$, where $\hat{b}_{l,k}$ is the Fourier transform of $\hat{\tilde b}_{l,j}$. Its dispersion relation reads $\omega_k=\omega_r+2\xi\cos k$, which shows a finite bandwidth $4\xi$ centered at $\omega_r$. The spectral density is $J(\omega)={g^2\over 2\pi\xi^2}\sqrt{4\xi^2-(\omega-\omega_r)^2}$. Figure \ref{carrf} shows the energy spectrum of the total system formed by either the superconductor qubit or the LC resonator and its coupled resonator array and $Z$ in different $g$. It is found that, with increasing $g$, a bound state is present in the energy spectrum. The formation of the bound state results in an abrupt jump of $Z$, which is equal to the long-time value of $|u(t)|$, from 0 to a positive value. It is readily obtained from Eqs. \eqref{17} and \eqref{32} that, as long as $Z$ is larger than 0, the quantum superiority is recovered for both of the discrete- and continuous-variable quantum teleportation schemes even till the long-time steady state.

The effects of the non-Markovian dynamics and the bound state have been observed in recent experiments \cite{yu2018experimental,lu2020observing,liu2017quantum,krinner2018spontaneous,Kwon2022}, which provide a strong support in the experimental realization of our noise-mitigation scheme.

\section{Conclusion}\label{Dsccon}
In summary, we have proposed a noise-mitigation mechanism for quantum teleportation to be universal for both the discrete- and continuous-variable quantum channels, under which the the quantum superiority in the teleportation fidelity is retrieved. This is in sharp contrast to the result under the Born-Markovian approximation, where the fidelity exponentially decays to, or even worse than, its classical limit. Our analysis reveals that it is due to the constructive interplay between the non-Markovian effect and the bound state of the total system consisting of the involved systems of the quantum channel and their reservoirs: The bound state supplies the intrinsic ability and the non-Markovian effect supplies the dynamical way to achieve the good performance. Efficiently overcoming the decoherence obstacle in realizing quantum teleportation, our result provides an experimentally feasible strategy to realize high-fidelity teleportation in practice.

\section*{Acknowledgments}
This work is supported by the National Natural Science Foundation of China (Grants No. 12275109 and No. 12247101), the Innovation Program for Quantum Science and Technology (Grant No. 2023ZD0300904) of China, and the Supercomputing Center of Lanzhou University.
\bibliography{cite}

\begin{thebibliography}{82}%
\makeatletter
\providecommand \@ifxundefined [1]{%
 \@ifx{#1\undefined}
}%
\providecommand \@ifnum [1]{%
 \ifnum #1\expandafter \@firstoftwo
 \else \expandafter \@secondoftwo
 \fi
}%
\providecommand \@ifx [1]{%
 \ifx #1\expandafter \@firstoftwo
 \else \expandafter \@secondoftwo
 \fi
}%
\providecommand \natexlab [1]{#1}%
\providecommand \enquote  [1]{``#1''}%
\providecommand \bibnamefont  [1]{#1}%
\providecommand \bibfnamefont [1]{#1}%
\providecommand \citenamefont [1]{#1}%
\providecommand \href@noop [0]{\@secondoftwo}%
\providecommand \href [0]{\begingroup \@sanitize@url \@href}%
\providecommand \@href[1]{\@@startlink{#1}\@@href}%
\providecommand \@@href[1]{\endgroup#1\@@endlink}%
\providecommand \@sanitize@url [0]{\catcode `\\12\catcode `\$12\catcode
  `\&12\catcode `\#12\catcode `\^12\catcode `\_12\catcode `\%12\relax}%
\providecommand \@@startlink[1]{}%
\providecommand \@@endlink[0]{}%
\providecommand \url  [0]{\begingroup\@sanitize@url \@url }%
\providecommand \@url [1]{\endgroup\@href {#1}{\urlprefix }}%
\providecommand \urlprefix  [0]{URL }%
\providecommand \Eprint [0]{\href }%
\providecommand \doibase [0]{https://doi.org/}%
\providecommand \selectlanguage [0]{\@gobble}%
\providecommand \bibinfo  [0]{\@secondoftwo}%
\providecommand \bibfield  [0]{\@secondoftwo}%
\providecommand \translation [1]{[#1]}%
\providecommand \BibitemOpen [0]{}%
\providecommand \bibitemStop [0]{}%
\providecommand \bibitemNoStop [0]{.\EOS\space}%
\providecommand \EOS [0]{\spacefactor3000\relax}%
\providecommand \BibitemShut  [1]{\csname bibitem#1\endcsname}%
\let\auto@bib@innerbib\@empty
\bibitem [{\citenamefont {Bennett}\ \emph {et~al.}(1993)\citenamefont
  {Bennett}, \citenamefont {Brassard}, \citenamefont {Cr\'epeau}, \citenamefont
  {Jozsa}, \citenamefont {Peres},\ and\ \citenamefont
  {Wootters}}]{PhysRevLett.70.1895}%
  \BibitemOpen
  \bibfield  {author} {\bibinfo {author} {\bibfnamefont {C.~H.}\ \bibnamefont
  {Bennett}}, \bibinfo {author} {\bibfnamefont {G.}~\bibnamefont {Brassard}},
  \bibinfo {author} {\bibfnamefont {C.}~\bibnamefont {Cr\'epeau}}, \bibinfo
  {author} {\bibfnamefont {R.}~\bibnamefont {Jozsa}}, \bibinfo {author}
  {\bibfnamefont {A.}~\bibnamefont {Peres}},\ and\ \bibinfo {author}
  {\bibfnamefont {W.~K.}\ \bibnamefont {Wootters}},\ }\bibfield  {title}
  {\bibinfo {title} {Teleporting an unknown quantum state via dual classical
  and {E}instein-{P}odolsky-{R}osen channels},\ }\href
  {https://doi.org/10.1103/PhysRevLett.70.1895} {\bibfield  {journal} {\bibinfo
   {journal} {Phys. Rev. Lett.}\ }\textbf {\bibinfo {volume} {70}},\ \bibinfo
  {pages} {1895} (\bibinfo {year} {1993})}\BibitemShut {NoStop}%
\bibitem [{\citenamefont {Pirandola}\ \emph {et~al.}(2015)\citenamefont
  {Pirandola}, \citenamefont {Eisert}, \citenamefont {Weedbrook}, \citenamefont
  {Furusawa},\ and\ \citenamefont {Braunstein}}]{Pirandola2015}%
  \BibitemOpen
  \bibfield  {author} {\bibinfo {author} {\bibfnamefont {S.}~\bibnamefont
  {Pirandola}}, \bibinfo {author} {\bibfnamefont {J.}~\bibnamefont {Eisert}},
  \bibinfo {author} {\bibfnamefont {C.}~\bibnamefont {Weedbrook}}, \bibinfo
  {author} {\bibfnamefont {A.}~\bibnamefont {Furusawa}},\ and\ \bibinfo
  {author} {\bibfnamefont {S.~L.}\ \bibnamefont {Braunstein}},\ }\bibfield
  {title} {\bibinfo {title} {Advances in quantum teleportation},\ }\href
  {https://doi.org/10.1038/nphoton.2015.154} {\bibfield  {journal} {\bibinfo
  {journal} {Nature Photonics}\ }\textbf {\bibinfo {volume} {9}},\ \bibinfo
  {pages} {641} (\bibinfo {year} {2015})}\BibitemShut {NoStop}%
\bibitem [{\citenamefont {Hu}\ \emph {et~al.}(2023)\citenamefont {Hu},
  \citenamefont {Guo}, \citenamefont {Liu}, \citenamefont {Li},\ and\
  \citenamefont {Guo}}]{HuXM2023}%
  \BibitemOpen
  \bibfield  {author} {\bibinfo {author} {\bibfnamefont {X.-M.}\ \bibnamefont
  {Hu}}, \bibinfo {author} {\bibfnamefont {Y.}~\bibnamefont {Guo}}, \bibinfo
  {author} {\bibfnamefont {B.-H.}\ \bibnamefont {Liu}}, \bibinfo {author}
  {\bibfnamefont {C.-F.}\ \bibnamefont {Li}},\ and\ \bibinfo {author}
  {\bibfnamefont {G.-C.}\ \bibnamefont {Guo}},\ }\bibfield  {title} {\bibinfo
  {title} {Progress in quantum teleportation},\ }\href
  {https://doi.org/10.1038/s42254-023-00588-x} {\bibfield  {journal} {\bibinfo
  {journal} {Nature Reviews Physics}\ }\textbf {\bibinfo {volume} {5}},\
  \bibinfo {pages} {339} (\bibinfo {year} {2023})}\BibitemShut {NoStop}%
\bibitem [{\citenamefont {Briegel}\ \emph {et~al.}(1998)\citenamefont
  {Briegel}, \citenamefont {D\"ur}, \citenamefont {Cirac},\ and\ \citenamefont
  {Zoller}}]{PhysRevLett.81.5932}%
  \BibitemOpen
  \bibfield  {author} {\bibinfo {author} {\bibfnamefont {H.-J.}\ \bibnamefont
  {Briegel}}, \bibinfo {author} {\bibfnamefont {W.}~\bibnamefont {D\"ur}},
  \bibinfo {author} {\bibfnamefont {J.~I.}\ \bibnamefont {Cirac}},\ and\
  \bibinfo {author} {\bibfnamefont {P.}~\bibnamefont {Zoller}},\ }\bibfield
  {title} {\bibinfo {title} {Quantum repeaters: The role of imperfect local
  operations in quantum communication},\ }\href
  {https://doi.org/10.1103/PhysRevLett.81.5932} {\bibfield  {journal} {\bibinfo
   {journal} {Phys. Rev. Lett.}\ }\textbf {\bibinfo {volume} {81}},\ \bibinfo
  {pages} {5932} (\bibinfo {year} {1998})}\BibitemShut {NoStop}%
\bibitem [{\citenamefont {Cacciapuoti}\ \emph {et~al.}(2020)\citenamefont
  {Cacciapuoti}, \citenamefont {Caleffi}, \citenamefont {Van~Meter},\ and\
  \citenamefont {Hanzo}}]{9023997}%
  \BibitemOpen
  \bibfield  {author} {\bibinfo {author} {\bibfnamefont {A.~S.}\ \bibnamefont
  {Cacciapuoti}}, \bibinfo {author} {\bibfnamefont {M.}~\bibnamefont
  {Caleffi}}, \bibinfo {author} {\bibfnamefont {R.}~\bibnamefont {Van~Meter}},\
  and\ \bibinfo {author} {\bibfnamefont {L.}~\bibnamefont {Hanzo}},\ }\bibfield
   {title} {\bibinfo {title} {When entanglement meets classical communications:
  Quantum teleportation for the quantum internet},\ }\href
  {https://doi.org/10.1109/TCOMM.2020.2978071} {\bibfield  {journal} {\bibinfo
  {journal} {IEEE Transactions on Communications}\ }\textbf {\bibinfo {volume}
  {68}},\ \bibinfo {pages} {3808} (\bibinfo {year} {2020})}\BibitemShut
  {NoStop}%
\bibitem [{\citenamefont {Brassard}\ \emph {et~al.}(1998)\citenamefont
  {Brassard}, \citenamefont {Braunstein},\ and\ \citenamefont
  {Cleve}}]{BRASSARD199843}%
  \BibitemOpen
  \bibfield  {author} {\bibinfo {author} {\bibfnamefont {G.}~\bibnamefont
  {Brassard}}, \bibinfo {author} {\bibfnamefont {S.~L.}\ \bibnamefont
  {Braunstein}},\ and\ \bibinfo {author} {\bibfnamefont {R.}~\bibnamefont
  {Cleve}},\ }\bibfield  {title} {\bibinfo {title} {Teleportation as a quantum
  computation},\ }\href
  {https://doi.org/https://doi.org/10.1016/S0167-2789(98)00043-8} {\bibfield
  {journal} {\bibinfo  {journal} {Physica D: Nonlinear Phenomena}\ }\textbf
  {\bibinfo {volume} {120}},\ \bibinfo {pages} {43} (\bibinfo {year}
  {1998})}\BibitemShut {NoStop}%
\bibitem [{\citenamefont {Gottesman}\ and\ \citenamefont
  {Chuang}(1999)}]{Gottesman1999}%
  \BibitemOpen
  \bibfield  {author} {\bibinfo {author} {\bibfnamefont {D.}~\bibnamefont
  {Gottesman}}\ and\ \bibinfo {author} {\bibfnamefont {I.~L.}\ \bibnamefont
  {Chuang}},\ }\bibfield  {title} {\bibinfo {title} {Demonstrating the
  viability of universal quantum computation using teleportation and
  single-qubit operations},\ }\href {https://doi.org/10.1038/46503} {\bibfield
  {journal} {\bibinfo  {journal} {Nature (London)}\ }\textbf {\bibinfo {volume}
  {402}},\ \bibinfo {pages} {390} (\bibinfo {year} {1999})}\BibitemShut
  {NoStop}%
\bibitem [{\citenamefont {Raussendorf}\ and\ \citenamefont
  {Briegel}(2001)}]{PhysRevLett.86.5188}%
  \BibitemOpen
  \bibfield  {author} {\bibinfo {author} {\bibfnamefont {R.}~\bibnamefont
  {Raussendorf}}\ and\ \bibinfo {author} {\bibfnamefont {H.~J.}\ \bibnamefont
  {Briegel}},\ }\bibfield  {title} {\bibinfo {title} {A one-way quantum
  computer},\ }\href {https://doi.org/10.1103/PhysRevLett.86.5188} {\bibfield
  {journal} {\bibinfo  {journal} {Phys. Rev. Lett.}\ }\textbf {\bibinfo
  {volume} {86}},\ \bibinfo {pages} {5188} (\bibinfo {year}
  {2001})}\BibitemShut {NoStop}%
\bibitem [{\citenamefont {Ishizaka}\ and\ \citenamefont
  {Hiroshima}(2008)}]{PhysRevLett.101.240501}%
  \BibitemOpen
  \bibfield  {author} {\bibinfo {author} {\bibfnamefont {S.}~\bibnamefont
  {Ishizaka}}\ and\ \bibinfo {author} {\bibfnamefont {T.}~\bibnamefont
  {Hiroshima}},\ }\bibfield  {title} {\bibinfo {title} {Asymptotic
  teleportation scheme as a universal programmable quantum processor},\ }\href
  {https://doi.org/10.1103/PhysRevLett.101.240501} {\bibfield  {journal}
  {\bibinfo  {journal} {Phys. Rev. Lett.}\ }\textbf {\bibinfo {volume} {101}},\
  \bibinfo {pages} {240501} (\bibinfo {year} {2008})}\BibitemShut {NoStop}%
\bibitem [{\citenamefont {Kimble}(2008)}]{Kimble2008}%
  \BibitemOpen
  \bibfield  {author} {\bibinfo {author} {\bibfnamefont {H.~J.}\ \bibnamefont
  {Kimble}},\ }\bibfield  {title} {\bibinfo {title} {The quantum internet},\
  }\href {https://doi.org/10.1038/nature07127} {\bibfield  {journal} {\bibinfo
  {journal} {Nature (London)}\ }\textbf {\bibinfo {volume} {453}},\ \bibinfo
  {pages} {1023} (\bibinfo {year} {2008})}\BibitemShut {NoStop}%
\bibitem [{\citenamefont {Lee}\ \emph {et~al.}(2020)\citenamefont {Lee},
  \citenamefont {Lee}, \citenamefont {Jeong},\ and\ \citenamefont
  {Park}}]{PhysRevLett.124.060501}%
  \BibitemOpen
  \bibfield  {author} {\bibinfo {author} {\bibfnamefont {S.~M.}\ \bibnamefont
  {Lee}}, \bibinfo {author} {\bibfnamefont {S.-W.}\ \bibnamefont {Lee}},
  \bibinfo {author} {\bibfnamefont {H.}~\bibnamefont {Jeong}},\ and\ \bibinfo
  {author} {\bibfnamefont {H.~S.}\ \bibnamefont {Park}},\ }\bibfield  {title}
  {\bibinfo {title} {Quantum teleportation of shared quantum secret},\ }\href
  {https://doi.org/10.1103/PhysRevLett.124.060501} {\bibfield  {journal}
  {\bibinfo  {journal} {Phys. Rev. Lett.}\ }\textbf {\bibinfo {volume} {124}},\
  \bibinfo {pages} {060501} (\bibinfo {year} {2020})}\BibitemShut {NoStop}%
\bibitem [{\citenamefont {Lloyd}\ \emph {et~al.}(2011)\citenamefont {Lloyd},
  \citenamefont {Maccone}, \citenamefont {Garcia-Patron}, \citenamefont
  {Giovannetti}, \citenamefont {Shikano}, \citenamefont {Pirandola},
  \citenamefont {Rozema}, \citenamefont {Darabi}, \citenamefont {Soudagar},
  \citenamefont {Shalm},\ and\ \citenamefont
  {Steinberg}}]{PhysRevLett.106.040403}%
  \BibitemOpen
  \bibfield  {author} {\bibinfo {author} {\bibfnamefont {S.}~\bibnamefont
  {Lloyd}}, \bibinfo {author} {\bibfnamefont {L.}~\bibnamefont {Maccone}},
  \bibinfo {author} {\bibfnamefont {R.}~\bibnamefont {Garcia-Patron}}, \bibinfo
  {author} {\bibfnamefont {V.}~\bibnamefont {Giovannetti}}, \bibinfo {author}
  {\bibfnamefont {Y.}~\bibnamefont {Shikano}}, \bibinfo {author} {\bibfnamefont
  {S.}~\bibnamefont {Pirandola}}, \bibinfo {author} {\bibfnamefont {L.~A.}\
  \bibnamefont {Rozema}}, \bibinfo {author} {\bibfnamefont {A.}~\bibnamefont
  {Darabi}}, \bibinfo {author} {\bibfnamefont {Y.}~\bibnamefont {Soudagar}},
  \bibinfo {author} {\bibfnamefont {L.~K.}\ \bibnamefont {Shalm}},\ and\
  \bibinfo {author} {\bibfnamefont {A.~M.}\ \bibnamefont {Steinberg}},\
  }\bibfield  {title} {\bibinfo {title} {Closed timelike curves via
  postselection: Theory and experimental test of consistency},\ }\href
  {https://doi.org/10.1103/PhysRevLett.106.040403} {\bibfield  {journal}
  {\bibinfo  {journal} {Phys. Rev. Lett.}\ }\textbf {\bibinfo {volume} {106}},\
  \bibinfo {pages} {040403} (\bibinfo {year} {2011})}\BibitemShut {NoStop}%
\bibitem [{\citenamefont {Lloyd}\ and\ \citenamefont
  {Preskill}(2014)}]{Lloyd2014}%
  \BibitemOpen
  \bibfield  {author} {\bibinfo {author} {\bibfnamefont {S.}~\bibnamefont
  {Lloyd}}\ and\ \bibinfo {author} {\bibfnamefont {J.}~\bibnamefont
  {Preskill}},\ }\bibfield  {title} {\bibinfo {title} {Unitarity of black hole
  evaporation in final-state projection models},\ }\href
  {https://doi.org/10.1007/JHEP08(2014)126} {\bibfield  {journal} {\bibinfo
  {journal} {Journal of High Energy Physics}\ }\textbf {\bibinfo {volume}
  {2014}},\ \bibinfo {pages} {126} (\bibinfo {year} {2014})}\BibitemShut
  {NoStop}%
\bibitem [{\citenamefont {Braunstein}\ and\ \citenamefont
  {Kimble}(1998)}]{PhysRevLett.80.869}%
  \BibitemOpen
  \bibfield  {author} {\bibinfo {author} {\bibfnamefont {S.~L.}\ \bibnamefont
  {Braunstein}}\ and\ \bibinfo {author} {\bibfnamefont {H.~J.}\ \bibnamefont
  {Kimble}},\ }\bibfield  {title} {\bibinfo {title} {Teleportation of
  continuous quantum variables},\ }\href
  {https://doi.org/10.1103/PhysRevLett.80.869} {\bibfield  {journal} {\bibinfo
  {journal} {Phys. Rev. Lett.}\ }\textbf {\bibinfo {volume} {80}},\ \bibinfo
  {pages} {869} (\bibinfo {year} {1998})}\BibitemShut {NoStop}%
\bibitem [{\citenamefont {Bouwmeester}\ \emph {et~al.}(1997)\citenamefont
  {Bouwmeester}, \citenamefont {Pan}, \citenamefont {Mattle}, \citenamefont
  {Eibl}, \citenamefont {Weinfurter},\ and\ \citenamefont
  {Zeilinger}}]{Bouwmeester1997}%
  \BibitemOpen
  \bibfield  {author} {\bibinfo {author} {\bibfnamefont {D.}~\bibnamefont
  {Bouwmeester}}, \bibinfo {author} {\bibfnamefont {J.-W.}\ \bibnamefont
  {Pan}}, \bibinfo {author} {\bibfnamefont {K.}~\bibnamefont {Mattle}},
  \bibinfo {author} {\bibfnamefont {M.}~\bibnamefont {Eibl}}, \bibinfo {author}
  {\bibfnamefont {H.}~\bibnamefont {Weinfurter}},\ and\ \bibinfo {author}
  {\bibfnamefont {A.}~\bibnamefont {Zeilinger}},\ }\bibfield  {title} {\bibinfo
  {title} {Experimental quantum teleportation},\ }\href
  {https://doi.org/10.1038/37539} {\bibfield  {journal} {\bibinfo  {journal}
  {Nature (London)}\ }\textbf {\bibinfo {volume} {390}},\ \bibinfo {pages}
  {575} (\bibinfo {year} {1997})}\BibitemShut {NoStop}%
\bibitem [{\citenamefont {Ma}\ \emph {et~al.}(2012)\citenamefont {Ma} \emph
  {et~al.}}]{ma2012quantum}%
  \BibitemOpen
  \bibfield  {author} {\bibinfo {author} {\bibfnamefont {X.-S.}\ \bibnamefont
  {Ma}} \emph {et~al.},\ }\bibfield  {title} {\bibinfo {title} {Quantum
  teleportation over 143 kilometres using active feed-forward},\ }\href
  {https://doi.org/10.1038/nature11472} {\bibfield  {journal} {\bibinfo
  {journal} {Nature (London)}\ }\textbf {\bibinfo {volume} {489}},\ \bibinfo
  {pages} {269} (\bibinfo {year} {2012})}\BibitemShut {NoStop}%
\bibitem [{\citenamefont {Luo}\ \emph {et~al.}(2019)\citenamefont {Luo},
  \citenamefont {Zhong}, \citenamefont {Erhard}, \citenamefont {Wang},
  \citenamefont {Peng}, \citenamefont {Krenn}, \citenamefont {Jiang},
  \citenamefont {Li}, \citenamefont {Liu}, \citenamefont {Lu}, \citenamefont
  {Zeilinger},\ and\ \citenamefont {Pan}}]{PhysRevLett.123.070505}%
  \BibitemOpen
  \bibfield  {author} {\bibinfo {author} {\bibfnamefont {Y.-H.}\ \bibnamefont
  {Luo}}, \bibinfo {author} {\bibfnamefont {H.-S.}\ \bibnamefont {Zhong}},
  \bibinfo {author} {\bibfnamefont {M.}~\bibnamefont {Erhard}}, \bibinfo
  {author} {\bibfnamefont {X.-L.}\ \bibnamefont {Wang}}, \bibinfo {author}
  {\bibfnamefont {L.-C.}\ \bibnamefont {Peng}}, \bibinfo {author}
  {\bibfnamefont {M.}~\bibnamefont {Krenn}}, \bibinfo {author} {\bibfnamefont
  {X.}~\bibnamefont {Jiang}}, \bibinfo {author} {\bibfnamefont
  {L.}~\bibnamefont {Li}}, \bibinfo {author} {\bibfnamefont {N.-L.}\
  \bibnamefont {Liu}}, \bibinfo {author} {\bibfnamefont {C.-Y.}\ \bibnamefont
  {Lu}}, \bibinfo {author} {\bibfnamefont {A.}~\bibnamefont {Zeilinger}},\ and\
  \bibinfo {author} {\bibfnamefont {J.-W.}\ \bibnamefont {Pan}},\ }\bibfield
  {title} {\bibinfo {title} {Quantum teleportation in high dimensions},\ }\href
  {https://doi.org/10.1103/PhysRevLett.123.070505} {\bibfield  {journal}
  {\bibinfo  {journal} {Phys. Rev. Lett.}\ }\textbf {\bibinfo {volume} {123}},\
  \bibinfo {pages} {070505} (\bibinfo {year} {2019})}\BibitemShut {NoStop}%
\bibitem [{\citenamefont {Furusawa}\ \emph {et~al.}(1998)\citenamefont
  {Furusawa}, \citenamefont {Sørensen}, \citenamefont {Braunstein},
  \citenamefont {Fuchs}, \citenamefont {Kimble},\ and\ \citenamefont
  {Polzik}}]{doi:10.1126/science.282.5389.706}%
  \BibitemOpen
  \bibfield  {author} {\bibinfo {author} {\bibfnamefont {A.}~\bibnamefont
  {Furusawa}}, \bibinfo {author} {\bibfnamefont {J.~L.}\ \bibnamefont
  {Sørensen}}, \bibinfo {author} {\bibfnamefont {S.~L.}\ \bibnamefont
  {Braunstein}}, \bibinfo {author} {\bibfnamefont {C.~A.}\ \bibnamefont
  {Fuchs}}, \bibinfo {author} {\bibfnamefont {H.~J.}\ \bibnamefont {Kimble}},\
  and\ \bibinfo {author} {\bibfnamefont {E.~S.}\ \bibnamefont {Polzik}},\
  }\bibfield  {title} {\bibinfo {title} {Unconditional quantum teleportation},\
  }\href {https://doi.org/10.1126/science.282.5389.706} {\bibfield  {journal}
  {\bibinfo  {journal} {Science}\ }\textbf {\bibinfo {volume} {282}},\ \bibinfo
  {pages} {706} (\bibinfo {year} {1998})}\BibitemShut {NoStop}%
\bibitem [{\citenamefont {Lee}\ \emph {et~al.}(2011)\citenamefont {Lee},
  \citenamefont {Benichi}, \citenamefont {Takeno}, \citenamefont {Takeda},
  \citenamefont {Webb}, \citenamefont {Huntington},\ and\ \citenamefont
  {Furusawa}}]{doi:10.1126/science.1201034}%
  \BibitemOpen
  \bibfield  {author} {\bibinfo {author} {\bibfnamefont {N.}~\bibnamefont
  {Lee}}, \bibinfo {author} {\bibfnamefont {H.}~\bibnamefont {Benichi}},
  \bibinfo {author} {\bibfnamefont {Y.}~\bibnamefont {Takeno}}, \bibinfo
  {author} {\bibfnamefont {S.}~\bibnamefont {Takeda}}, \bibinfo {author}
  {\bibfnamefont {J.}~\bibnamefont {Webb}}, \bibinfo {author} {\bibfnamefont
  {E.}~\bibnamefont {Huntington}},\ and\ \bibinfo {author} {\bibfnamefont
  {A.}~\bibnamefont {Furusawa}},\ }\bibfield  {title} {\bibinfo {title}
  {Teleportation of nonclassical wave packets of light},\ }\href
  {https://doi.org/10.1126/science.1201034} {\bibfield  {journal} {\bibinfo
  {journal} {Science}\ }\textbf {\bibinfo {volume} {332}},\ \bibinfo {pages}
  {330} (\bibinfo {year} {2011})}\BibitemShut {NoStop}%
\bibitem [{\citenamefont {Georgescu}(2022)}]{Georgescu2022}%
  \BibitemOpen
  \bibfield  {author} {\bibinfo {author} {\bibfnamefont {I.}~\bibnamefont
  {Georgescu}},\ }\bibfield  {title} {\bibinfo {title} {25 years of
  experimental quantum teleportation},\ }\href
  {https://doi.org/10.1038/s42254-022-00530-7} {\bibfield  {journal} {\bibinfo
  {journal} {Nature Reviews Physics}\ }\textbf {\bibinfo {volume} {4}},\
  \bibinfo {pages} {695} (\bibinfo {year} {2022})}\BibitemShut {NoStop}%
\bibitem [{\citenamefont {Ren}\ \emph {et~al.}(2017)\citenamefont {Ren} \emph
  {et~al.}}]{ren2017ground}%
  \BibitemOpen
  \bibfield  {author} {\bibinfo {author} {\bibfnamefont {J.-G.}\ \bibnamefont
  {Ren}} \emph {et~al.},\ }\bibfield  {title} {\bibinfo {title}
  {Ground-to-satellite quantum teleportation},\ }\href
  {https://doi.org/10.1038/nature23675} {\bibfield  {journal} {\bibinfo
  {journal} {Nature (London)}\ }\textbf {\bibinfo {volume} {549}},\ \bibinfo
  {pages} {70} (\bibinfo {year} {2017})}\BibitemShut {NoStop}%
\bibitem [{\citenamefont {Preskill}(2018)}]{Preskill2018quantumcomputingin}%
  \BibitemOpen
  \bibfield  {author} {\bibinfo {author} {\bibfnamefont {J.}~\bibnamefont
  {Preskill}},\ }\bibfield  {title} {\bibinfo {title} {Quantum {C}omputing in
  the {NISQ} era and beyond},\ }\href
  {https://doi.org/10.22331/q-2018-08-06-79} {\bibfield  {journal} {\bibinfo
  {journal} {{Quantum}}\ }\textbf {\bibinfo {volume} {2}},\ \bibinfo {pages}
  {79} (\bibinfo {year} {2018})}\BibitemShut {NoStop}%
\bibitem [{\citenamefont {Chen}\ \emph {et~al.}(2023)\citenamefont {Chen},
  \citenamefont {Cotler}, \citenamefont {Huang},\ and\ \citenamefont
  {Li}}]{Chen2023}%
  \BibitemOpen
  \bibfield  {author} {\bibinfo {author} {\bibfnamefont {S.}~\bibnamefont
  {Chen}}, \bibinfo {author} {\bibfnamefont {J.}~\bibnamefont {Cotler}},
  \bibinfo {author} {\bibfnamefont {H.-Y.}\ \bibnamefont {Huang}},\ and\
  \bibinfo {author} {\bibfnamefont {J.}~\bibnamefont {Li}},\ }\bibfield
  {title} {\bibinfo {title} {The complexity of {NISQ}},\ }\href
  {https://doi.org/10.1038/s41467-023-41217-6} {\bibfield  {journal} {\bibinfo
  {journal} {Nature Communications}\ }\textbf {\bibinfo {volume} {14}},\
  \bibinfo {pages} {6001} (\bibinfo {year} {2023})}\BibitemShut {NoStop}%
\bibitem [{\citenamefont {Jiao}\ \emph {et~al.}(2024)\citenamefont {Jiao},
  \citenamefont {Wu}, \citenamefont {Bai},\ and\ \citenamefont
  {An}}]{https://doi.org/10.1002/qute.202300218}%
  \BibitemOpen
  \bibfield  {author} {\bibinfo {author} {\bibfnamefont {L.}~\bibnamefont
  {Jiao}}, \bibinfo {author} {\bibfnamefont {W.}~\bibnamefont {Wu}}, \bibinfo
  {author} {\bibfnamefont {S.-Y.}\ \bibnamefont {Bai}},\ and\ \bibinfo {author}
  {\bibfnamefont {J.-H.}\ \bibnamefont {An}},\ }\bibfield  {title} {\bibinfo
  {title} {Quantum metrology in the noisy intermediate-scale quantum era},\
  }\href {https://doi.org/https://doi.org/10.1002/qute.202300218} {\bibfield
  {journal} {\bibinfo  {journal} {Advanced Quantum Technologies}\ ,\ \bibinfo
  {pages} {2300218}} (\bibinfo {year} {2024})}\BibitemShut {NoStop}%
\bibitem [{\citenamefont {Bennett}\ \emph {et~al.}(1996)\citenamefont
  {Bennett}, \citenamefont {Brassard}, \citenamefont {Popescu}, \citenamefont
  {Schumacher}, \citenamefont {Smolin},\ and\ \citenamefont
  {Wootters}}]{PhysRevLett.76.722}%
  \BibitemOpen
  \bibfield  {author} {\bibinfo {author} {\bibfnamefont {C.~H.}\ \bibnamefont
  {Bennett}}, \bibinfo {author} {\bibfnamefont {G.}~\bibnamefont {Brassard}},
  \bibinfo {author} {\bibfnamefont {S.}~\bibnamefont {Popescu}}, \bibinfo
  {author} {\bibfnamefont {B.}~\bibnamefont {Schumacher}}, \bibinfo {author}
  {\bibfnamefont {J.~A.}\ \bibnamefont {Smolin}},\ and\ \bibinfo {author}
  {\bibfnamefont {W.~K.}\ \bibnamefont {Wootters}},\ }\bibfield  {title}
  {\bibinfo {title} {Purification of noisy entanglement and faithful
  teleportation via noisy channels},\ }\href
  {https://doi.org/10.1103/PhysRevLett.76.722} {\bibfield  {journal} {\bibinfo
  {journal} {Phys. Rev. Lett.}\ }\textbf {\bibinfo {volume} {76}},\ \bibinfo
  {pages} {722} (\bibinfo {year} {1996})}\BibitemShut {NoStop}%
\bibitem [{\citenamefont {Lee}\ and\ \citenamefont
  {Kim}(2000)}]{lee2000entanglement}%
  \BibitemOpen
  \bibfield  {author} {\bibinfo {author} {\bibfnamefont {J.}~\bibnamefont
  {Lee}}\ and\ \bibinfo {author} {\bibfnamefont {M.~S.}\ \bibnamefont {Kim}},\
  }\bibfield  {title} {\bibinfo {title} {Entanglement teleportation via
  {W}erner states},\ }\href {https://doi.org/10.1103/PhysRevLett.84.4236}
  {\bibfield  {journal} {\bibinfo  {journal} {Phys. Rev. Lett.}\ }\textbf
  {\bibinfo {volume} {84}},\ \bibinfo {pages} {4236} (\bibinfo {year}
  {2000})}\BibitemShut {NoStop}%
\bibitem [{\citenamefont {Oh}\ \emph {et~al.}(2002)\citenamefont {Oh},
  \citenamefont {Lee},\ and\ \citenamefont {Lee}}]{oh2002fidelity}%
  \BibitemOpen
  \bibfield  {author} {\bibinfo {author} {\bibfnamefont {S.}~\bibnamefont
  {Oh}}, \bibinfo {author} {\bibfnamefont {S.}~\bibnamefont {Lee}},\ and\
  \bibinfo {author} {\bibfnamefont {H.-w.}\ \bibnamefont {Lee}},\ }\bibfield
  {title} {\bibinfo {title} {Fidelity of quantum teleportation through noisy
  channels},\ }\href {https://doi.org/10.1103/PhysRevA.66.022316} {\bibfield
  {journal} {\bibinfo  {journal} {Phys. Rev. A}\ }\textbf {\bibinfo {volume}
  {66}},\ \bibinfo {pages} {022316} (\bibinfo {year} {2002})}\BibitemShut
  {NoStop}%
\bibitem [{\citenamefont {Jung}\ \emph {et~al.}(2008)\citenamefont {Jung},
  \citenamefont {Hwang}, \citenamefont {Ju}, \citenamefont {Kim}, \citenamefont
  {Yoo}, \citenamefont {Kim}, \citenamefont {Park}, \citenamefont {Son},
  \citenamefont {Tamaryan},\ and\ \citenamefont {Cha}}]{jung2008greenberger}%
  \BibitemOpen
  \bibfield  {author} {\bibinfo {author} {\bibfnamefont {E.}~\bibnamefont
  {Jung}}, \bibinfo {author} {\bibfnamefont {M.-R.}\ \bibnamefont {Hwang}},
  \bibinfo {author} {\bibfnamefont {Y.~H.}\ \bibnamefont {Ju}}, \bibinfo
  {author} {\bibfnamefont {M.-S.}\ \bibnamefont {Kim}}, \bibinfo {author}
  {\bibfnamefont {S.-K.}\ \bibnamefont {Yoo}}, \bibinfo {author} {\bibfnamefont
  {H.}~\bibnamefont {Kim}}, \bibinfo {author} {\bibfnamefont {D.}~\bibnamefont
  {Park}}, \bibinfo {author} {\bibfnamefont {J.-W.}\ \bibnamefont {Son}},
  \bibinfo {author} {\bibfnamefont {S.}~\bibnamefont {Tamaryan}},\ and\
  \bibinfo {author} {\bibfnamefont {S.-K.}\ \bibnamefont {Cha}},\ }\bibfield
  {title} {\bibinfo {title} {{G}reenberger-{H}orne-{Z}eilinger versus {W}
  states: Quantum teleportation through noisy channels},\ }\href
  {https://doi.org/10.1103/PhysRevA.78.012312} {\bibfield  {journal} {\bibinfo
  {journal} {Phys. Rev. A}\ }\textbf {\bibinfo {volume} {78}},\ \bibinfo
  {pages} {012312} (\bibinfo {year} {2008})}\BibitemShut {NoStop}%
\bibitem [{\citenamefont {Rao}\ \emph {et~al.}(2008)\citenamefont {Rao},
  \citenamefont {Panigrahi},\ and\ \citenamefont
  {Mitra}}]{rao2008teleportation}%
  \BibitemOpen
  \bibfield  {author} {\bibinfo {author} {\bibfnamefont {D.~D.~B.}\
  \bibnamefont {Rao}}, \bibinfo {author} {\bibfnamefont {P.~K.}\ \bibnamefont
  {Panigrahi}},\ and\ \bibinfo {author} {\bibfnamefont {C.}~\bibnamefont
  {Mitra}},\ }\bibfield  {title} {\bibinfo {title} {Teleportation in the
  presence of common bath decoherence at the transmitting station},\ }\href
  {https://doi.org/10.1103/PhysRevA.78.022336} {\bibfield  {journal} {\bibinfo
  {journal} {Phys. Rev. A}\ }\textbf {\bibinfo {volume} {78}},\ \bibinfo
  {pages} {022336} (\bibinfo {year} {2008})}\BibitemShut {NoStop}%
\bibitem [{\citenamefont {Yeo}\ \emph {et~al.}(2009)\citenamefont {Yeo},
  \citenamefont {Kho},\ and\ \citenamefont {Wang}}]{yeo2009effects}%
  \BibitemOpen
  \bibfield  {author} {\bibinfo {author} {\bibfnamefont {Y.}~\bibnamefont
  {Yeo}}, \bibinfo {author} {\bibfnamefont {Z.-W.}\ \bibnamefont {Kho}},\ and\
  \bibinfo {author} {\bibfnamefont {L.}~\bibnamefont {Wang}},\ }\bibfield
  {title} {\bibinfo {title} {Effects of {P}auli channels and noisy quantum
  operations on standard teleportation},\ }\href
  {https://doi.org/10.1209/0295-5075/86/40009} {\bibfield  {journal} {\bibinfo
  {journal} {EPL (Europhysics Letters)}\ }\textbf {\bibinfo {volume} {86}},\
  \bibinfo {pages} {40009} (\bibinfo {year} {2009})}\BibitemShut {NoStop}%
\bibitem [{\citenamefont {Bowen}\ and\ \citenamefont
  {Bose}(2001)}]{PhysRevLett.87.267901}%
  \BibitemOpen
  \bibfield  {author} {\bibinfo {author} {\bibfnamefont {G.}~\bibnamefont
  {Bowen}}\ and\ \bibinfo {author} {\bibfnamefont {S.}~\bibnamefont {Bose}},\
  }\bibfield  {title} {\bibinfo {title} {Teleportation as a depolarizing
  quantum channel, relative entropy, and classical capacity},\ }\href
  {https://doi.org/10.1103/PhysRevLett.87.267901} {\bibfield  {journal}
  {\bibinfo  {journal} {Phys. Rev. Lett.}\ }\textbf {\bibinfo {volume} {87}},\
  \bibinfo {pages} {267901} (\bibinfo {year} {2001})}\BibitemShut {NoStop}%
\bibitem [{\citenamefont {Taketani}\ \emph {et~al.}(2012)\citenamefont
  {Taketani}, \citenamefont {de~Melo},\ and\ \citenamefont
  {de~Matos~Filho}}]{PhysRevA.85.020301}%
  \BibitemOpen
  \bibfield  {author} {\bibinfo {author} {\bibfnamefont {B.~G.}\ \bibnamefont
  {Taketani}}, \bibinfo {author} {\bibfnamefont {F.}~\bibnamefont {de~Melo}},\
  and\ \bibinfo {author} {\bibfnamefont {R.~L.}\ \bibnamefont
  {de~Matos~Filho}},\ }\bibfield  {title} {\bibinfo {title} {Optimal
  teleportation with a noisy source},\ }\href
  {https://doi.org/10.1103/PhysRevA.85.020301} {\bibfield  {journal} {\bibinfo
  {journal} {Phys. Rev. A}\ }\textbf {\bibinfo {volume} {85}},\ \bibinfo
  {pages} {020301(R)} (\bibinfo {year} {2012})}\BibitemShut {NoStop}%
\bibitem [{\citenamefont {Knoll}\ \emph {et~al.}(2014)\citenamefont {Knoll},
  \citenamefont {Schmiegelow},\ and\ \citenamefont
  {Larotonda}}]{PhysRevA.90.042332}%
  \BibitemOpen
  \bibfield  {author} {\bibinfo {author} {\bibfnamefont {L.~T.}\ \bibnamefont
  {Knoll}}, \bibinfo {author} {\bibfnamefont {C.~T.}\ \bibnamefont
  {Schmiegelow}},\ and\ \bibinfo {author} {\bibfnamefont {M.~A.}\ \bibnamefont
  {Larotonda}},\ }\bibfield  {title} {\bibinfo {title} {Noisy quantum
  teleportation: An experimental study on the influence of local
  environments},\ }\href {https://doi.org/10.1103/PhysRevA.90.042332}
  {\bibfield  {journal} {\bibinfo  {journal} {Phys. Rev. A}\ }\textbf {\bibinfo
  {volume} {90}},\ \bibinfo {pages} {042332} (\bibinfo {year}
  {2014})}\BibitemShut {NoStop}%
\bibitem [{\citenamefont {Fortes}\ and\ \citenamefont
  {Rigolin}(2015)}]{PhysRevA.92.012338}%
  \BibitemOpen
  \bibfield  {author} {\bibinfo {author} {\bibfnamefont {R.}~\bibnamefont
  {Fortes}}\ and\ \bibinfo {author} {\bibfnamefont {G.}~\bibnamefont
  {Rigolin}},\ }\bibfield  {title} {\bibinfo {title} {Fighting noise with noise
  in realistic quantum teleportation},\ }\href
  {https://doi.org/10.1103/PhysRevA.92.012338} {\bibfield  {journal} {\bibinfo
  {journal} {Phys. Rev. A}\ }\textbf {\bibinfo {volume} {92}},\ \bibinfo
  {pages} {012338} (\bibinfo {year} {2015})}\BibitemShut {NoStop}%
\bibitem [{\citenamefont {Guo}\ \emph {et~al.}(2020)\citenamefont {Guo},
  \citenamefont {Tian}, \citenamefont {Zeng},\ and\ \citenamefont
  {Chen}}]{Guo2020}%
  \BibitemOpen
  \bibfield  {author} {\bibinfo {author} {\bibfnamefont {Y.-n.}\ \bibnamefont
  {Guo}}, \bibinfo {author} {\bibfnamefont {Q.-l.}\ \bibnamefont {Tian}},
  \bibinfo {author} {\bibfnamefont {K.}~\bibnamefont {Zeng}},\ and\ \bibinfo
  {author} {\bibfnamefont {P.-x.}\ \bibnamefont {Chen}},\ }\bibfield  {title}
  {\bibinfo {title} {Fidelity of quantum teleportation in correlated quantum
  channels},\ }\href {https://doi.org/10.1007/s11128-020-02675-9} {\bibfield
  {journal} {\bibinfo  {journal} {Quantum Information Processing}\ }\textbf
  {\bibinfo {volume} {19}},\ \bibinfo {pages} {182} (\bibinfo {year}
  {2020})}\BibitemShut {NoStop}%
\bibitem [{\citenamefont {Lee}\ \emph {et~al.}(2021)\citenamefont {Lee},
  \citenamefont {Im}, \citenamefont {Kim}, \citenamefont {Nha},\ and\
  \citenamefont {Kim}}]{PhysRevResearch.3.033119}%
  \BibitemOpen
  \bibfield  {author} {\bibinfo {author} {\bibfnamefont {S.-W.}\ \bibnamefont
  {Lee}}, \bibinfo {author} {\bibfnamefont {D.-G.}\ \bibnamefont {Im}},
  \bibinfo {author} {\bibfnamefont {Y.-H.}\ \bibnamefont {Kim}}, \bibinfo
  {author} {\bibfnamefont {H.}~\bibnamefont {Nha}},\ and\ \bibinfo {author}
  {\bibfnamefont {M.~S.}\ \bibnamefont {Kim}},\ }\bibfield  {title} {\bibinfo
  {title} {Quantum teleportation is a reversal of quantum measurement},\ }\href
  {https://doi.org/10.1103/PhysRevResearch.3.033119} {\bibfield  {journal}
  {\bibinfo  {journal} {Phys. Rev. Res.}\ }\textbf {\bibinfo {volume} {3}},\
  \bibinfo {pages} {033119} (\bibinfo {year} {2021})}\BibitemShut {NoStop}%
\bibitem [{\citenamefont {Im}\ \emph {et~al.}(2021)\citenamefont {Im},
  \citenamefont {Lee}, \citenamefont {Kim}, \citenamefont {Nha}, \citenamefont
  {Kim}, \citenamefont {Lee},\ and\ \citenamefont {Kim}}]{Im2021}%
  \BibitemOpen
  \bibfield  {author} {\bibinfo {author} {\bibfnamefont {D.-G.}\ \bibnamefont
  {Im}}, \bibinfo {author} {\bibfnamefont {C.-H.}\ \bibnamefont {Lee}},
  \bibinfo {author} {\bibfnamefont {Y.}~\bibnamefont {Kim}}, \bibinfo {author}
  {\bibfnamefont {H.}~\bibnamefont {Nha}}, \bibinfo {author} {\bibfnamefont
  {M.~S.}\ \bibnamefont {Kim}}, \bibinfo {author} {\bibfnamefont {S.-W.}\
  \bibnamefont {Lee}},\ and\ \bibinfo {author} {\bibfnamefont {Y.-H.}\
  \bibnamefont {Kim}},\ }\bibfield  {title} {\bibinfo {title} {Optimal
  teleportation via noisy quantum channels without additional qubit
  resources},\ }\href {https://doi.org/10.1038/s41534-021-00426-x} {\bibfield
  {journal} {\bibinfo  {journal} {npj Quantum Information}\ }\textbf {\bibinfo
  {volume} {7}},\ \bibinfo {pages} {86} (\bibinfo {year} {2021})}\BibitemShut
  {NoStop}%
\bibitem [{\citenamefont {Luo}\ \emph {et~al.}(2021)\citenamefont {Luo},
  \citenamefont {Chen}, \citenamefont {Erhard}, \citenamefont {Zhong},
  \citenamefont {Wu}, \citenamefont {Tang}, \citenamefont {Zhao}, \citenamefont
  {Wang}, \citenamefont {Fujii}, \citenamefont {Li}, \citenamefont {Liu},
  \citenamefont {Nemoto}, \citenamefont {Munro}, \citenamefont {Lu},
  \citenamefont {Zeilinger},\ and\ \citenamefont
  {Pan}}]{doi:10.1073/pnas.2026250118}%
  \BibitemOpen
  \bibfield  {author} {\bibinfo {author} {\bibfnamefont {Y.-H.}\ \bibnamefont
  {Luo}}, \bibinfo {author} {\bibfnamefont {M.-C.}\ \bibnamefont {Chen}},
  \bibinfo {author} {\bibfnamefont {M.}~\bibnamefont {Erhard}}, \bibinfo
  {author} {\bibfnamefont {H.-S.}\ \bibnamefont {Zhong}}, \bibinfo {author}
  {\bibfnamefont {D.}~\bibnamefont {Wu}}, \bibinfo {author} {\bibfnamefont
  {H.-Y.}\ \bibnamefont {Tang}}, \bibinfo {author} {\bibfnamefont
  {Q.}~\bibnamefont {Zhao}}, \bibinfo {author} {\bibfnamefont {X.-L.}\
  \bibnamefont {Wang}}, \bibinfo {author} {\bibfnamefont {K.}~\bibnamefont
  {Fujii}}, \bibinfo {author} {\bibfnamefont {L.}~\bibnamefont {Li}}, \bibinfo
  {author} {\bibfnamefont {N.-L.}\ \bibnamefont {Liu}}, \bibinfo {author}
  {\bibfnamefont {K.}~\bibnamefont {Nemoto}}, \bibinfo {author} {\bibfnamefont
  {W.~J.}\ \bibnamefont {Munro}}, \bibinfo {author} {\bibfnamefont {C.-Y.}\
  \bibnamefont {Lu}}, \bibinfo {author} {\bibfnamefont {A.}~\bibnamefont
  {Zeilinger}},\ and\ \bibinfo {author} {\bibfnamefont {J.-W.}\ \bibnamefont
  {Pan}},\ }\bibfield  {title} {\bibinfo {title} {Quantum teleportation of
  physical qubits into logical code spaces},\ }\href
  {https://doi.org/10.1073/pnas.2026250118} {\bibfield  {journal} {\bibinfo
  {journal} {Proceedings of the National Academy of Sciences}\ }\textbf
  {\bibinfo {volume} {118}},\ \bibinfo {pages} {e2026250118} (\bibinfo {year}
  {2021})}\BibitemShut {NoStop}%
\bibitem [{\citenamefont {Harraz}\ \emph {et~al.}(2022)\citenamefont {Harraz},
  \citenamefont {Cong},\ and\ \citenamefont {Nieto}}]{Harraz2022}%
  \BibitemOpen
  \bibfield  {author} {\bibinfo {author} {\bibfnamefont {S.}~\bibnamefont
  {Harraz}}, \bibinfo {author} {\bibfnamefont {S.}~\bibnamefont {Cong}},\ and\
  \bibinfo {author} {\bibfnamefont {J.~J.}\ \bibnamefont {Nieto}},\ }\bibfield
  {title} {\bibinfo {title} {Enhancing quantum teleportation fidelity under
  decoherence via weak measurement with flips},\ }\href
  {https://doi.org/10.1140/epjqt/s40507-022-00134-1} {\bibfield  {journal}
  {\bibinfo  {journal} {EPJ Quantum Technology}\ }\textbf {\bibinfo {volume}
  {9}},\ \bibinfo {pages} {15} (\bibinfo {year} {2022})}\BibitemShut {NoStop}%
\bibitem [{\citenamefont {Seida}\ \emph {et~al.}(2022)\citenamefont {Seida},
  \citenamefont {Seddik}, \citenamefont {Hassouni},\ and\ \citenamefont
  {Allati}}]{SEIDA2022128115}%
  \BibitemOpen
  \bibfield  {author} {\bibinfo {author} {\bibfnamefont {C.}~\bibnamefont
  {Seida}}, \bibinfo {author} {\bibfnamefont {S.}~\bibnamefont {Seddik}},
  \bibinfo {author} {\bibfnamefont {Y.}~\bibnamefont {Hassouni}},\ and\
  \bibinfo {author} {\bibfnamefont {A.~E.}\ \bibnamefont {Allati}},\ }\bibfield
   {title} {\bibinfo {title} {Memory effects on bidirectional teleportation},\
  }\href {https://doi.org/https://doi.org/10.1016/j.physa.2022.128115}
  {\bibfield  {journal} {\bibinfo  {journal} {Physica A: Statistical Mechanics
  and its Applications}\ }\textbf {\bibinfo {volume} {606}},\ \bibinfo {pages}
  {128115} (\bibinfo {year} {2022})}\BibitemShut {NoStop}%
\bibitem [{\citenamefont {Zhao}\ \emph {et~al.}(2023)\citenamefont {Zhao},
  \citenamefont {Jeng}, \citenamefont {Conlon}, \citenamefont {Tserkis},
  \citenamefont {Shajilal}, \citenamefont {Liu}, \citenamefont {Ralph},
  \citenamefont {Assad},\ and\ \citenamefont {Lam}}]{Zhao2023}%
  \BibitemOpen
  \bibfield  {author} {\bibinfo {author} {\bibfnamefont {J.}~\bibnamefont
  {Zhao}}, \bibinfo {author} {\bibfnamefont {H.}~\bibnamefont {Jeng}}, \bibinfo
  {author} {\bibfnamefont {L.~O.}\ \bibnamefont {Conlon}}, \bibinfo {author}
  {\bibfnamefont {S.}~\bibnamefont {Tserkis}}, \bibinfo {author} {\bibfnamefont
  {B.}~\bibnamefont {Shajilal}}, \bibinfo {author} {\bibfnamefont
  {K.}~\bibnamefont {Liu}}, \bibinfo {author} {\bibfnamefont {T.~C.}\
  \bibnamefont {Ralph}}, \bibinfo {author} {\bibfnamefont {S.~M.}\ \bibnamefont
  {Assad}},\ and\ \bibinfo {author} {\bibfnamefont {P.~K.}\ \bibnamefont
  {Lam}},\ }\bibfield  {title} {\bibinfo {title} {Enhancing quantum
  teleportation efficacy with noiseless linear amplification},\ }\href
  {https://doi.org/10.1038/s41467-023-40438-z} {\bibfield  {journal} {\bibinfo
  {journal} {Nature Communications}\ }\textbf {\bibinfo {volume} {14}},\
  \bibinfo {pages} {4745} (\bibinfo {year} {2023})}\BibitemShut {NoStop}%
\bibitem [{\citenamefont {Roszak}\ and\ \citenamefont
  {Korbicz}(2023)}]{Roszak2023purifying}%
  \BibitemOpen
  \bibfield  {author} {\bibinfo {author} {\bibfnamefont {K.}~\bibnamefont
  {Roszak}}\ and\ \bibinfo {author} {\bibfnamefont {J.~K.}\ \bibnamefont
  {Korbicz}},\ }\bibfield  {title} {\bibinfo {title} {Purifying
  teleportation},\ }\href {https://doi.org/10.22331/q-2023-02-16-923}
  {\bibfield  {journal} {\bibinfo  {journal} {{Quantum}}\ }\textbf {\bibinfo
  {volume} {7}},\ \bibinfo {pages} {923} (\bibinfo {year} {2023})}\BibitemShut
  {NoStop}%
\bibitem [{\citenamefont {{Google Quantum AI}}\ and\ \citenamefont
  {Collaborators}(2023)}]{Hoke2023}%
  \BibitemOpen
  \bibfield  {author} {\bibinfo {author} {\bibnamefont {{Google Quantum AI}}}\
  and\ \bibinfo {author} {\bibnamefont {Collaborators}},\ }\bibfield  {title}
  {\bibinfo {title} {Measurement-induced entanglement and teleportation on a
  noisy quantum processor},\ }\href
  {https://doi.org/10.1038/s41586-023-06505-7} {\bibfield  {journal} {\bibinfo
  {journal} {Nature (London)}\ }\textbf {\bibinfo {volume} {622}},\ \bibinfo
  {pages} {481} (\bibinfo {year} {2023})}\BibitemShut {NoStop}%
\bibitem [{\citenamefont {Rivas}\ \emph {et~al.}(2014)\citenamefont {Rivas},
  \citenamefont {Huelga},\ and\ \citenamefont {Plenio}}]{Rivas_2014}%
  \BibitemOpen
  \bibfield  {author} {\bibinfo {author} {\bibfnamefont {{\'{A}}.}~\bibnamefont
  {Rivas}}, \bibinfo {author} {\bibfnamefont {S.~F.}\ \bibnamefont {Huelga}},\
  and\ \bibinfo {author} {\bibfnamefont {M.~B.}\ \bibnamefont {Plenio}},\
  }\bibfield  {title} {\bibinfo {title} {Quantum non-{M}arkovianity:
  {C}haracterization, quantification and detection},\ }\href
  {https://doi.org/10.1088/0034-4885/77/9/094001} {\bibfield  {journal}
  {\bibinfo  {journal} {Rep. Prog Phys}\ }\textbf {\bibinfo {volume} {77}},\
  \bibinfo {pages} {094001} (\bibinfo {year} {2014})}\BibitemShut {NoStop}%
\bibitem [{\citenamefont {Breuer}\ \emph {et~al.}(2016)\citenamefont {Breuer},
  \citenamefont {Laine}, \citenamefont {Piilo},\ and\ \citenamefont
  {Vacchini}}]{RevModPhys.88.021002}%
  \BibitemOpen
  \bibfield  {author} {\bibinfo {author} {\bibfnamefont {H.-P.}\ \bibnamefont
  {Breuer}}, \bibinfo {author} {\bibfnamefont {E.-M.}\ \bibnamefont {Laine}},
  \bibinfo {author} {\bibfnamefont {J.}~\bibnamefont {Piilo}},\ and\ \bibinfo
  {author} {\bibfnamefont {B.}~\bibnamefont {Vacchini}},\ }\bibfield  {title}
  {\bibinfo {title} {Colloquium: Non-{M}arkovian dynamics in open quantum
  systems},\ }\href {https://doi.org/10.1103/RevModPhys.88.021002} {\bibfield
  {journal} {\bibinfo  {journal} {Rev. Mod. Phys.}\ }\textbf {\bibinfo {volume}
  {88}},\ \bibinfo {pages} {021002} (\bibinfo {year} {2016})}\BibitemShut
  {NoStop}%
\bibitem [{\citenamefont {Li}\ \emph {et~al.}(2018)\citenamefont {Li},
  \citenamefont {Hall},\ and\ \citenamefont {Wiseman}}]{LI20181}%
  \BibitemOpen
  \bibfield  {author} {\bibinfo {author} {\bibfnamefont {L.}~\bibnamefont
  {Li}}, \bibinfo {author} {\bibfnamefont {M.~J.}\ \bibnamefont {Hall}},\ and\
  \bibinfo {author} {\bibfnamefont {H.~M.}\ \bibnamefont {Wiseman}},\
  }\bibfield  {title} {\bibinfo {title} {Concepts of quantum
  non-{M}arkovianity: A hierarchy},\ }\href
  {https://doi.org/https://doi.org/10.1016/j.physrep.2018.07.001} {\bibfield
  {journal} {\bibinfo  {journal} {Phys. Rep.}\ }\textbf {\bibinfo {volume}
  {759}},\ \bibinfo {pages} {1 } (\bibinfo {year} {2018})}\BibitemShut
  {NoStop}%
\bibitem [{\citenamefont {Bai}\ \emph {et~al.}(2021)\citenamefont {Bai},
  \citenamefont {Chen}, \citenamefont {Wu},\ and\ \citenamefont
  {An}}]{doi:10.1080/23746149.2020.1870559}%
  \BibitemOpen
  \bibfield  {author} {\bibinfo {author} {\bibfnamefont {S.-Y.}\ \bibnamefont
  {Bai}}, \bibinfo {author} {\bibfnamefont {C.}~\bibnamefont {Chen}}, \bibinfo
  {author} {\bibfnamefont {H.}~\bibnamefont {Wu}},\ and\ \bibinfo {author}
  {\bibfnamefont {J.-H.}\ \bibnamefont {An}},\ }\bibfield  {title} {\bibinfo
  {title} {Quantum control in open and periodically driven systems},\ }\href
  {https://doi.org/10.1080/23746149.2020.1870559} {\bibfield  {journal}
  {\bibinfo  {journal} {Advances in Physics: X}\ }\textbf {\bibinfo {volume}
  {6}},\ \bibinfo {pages} {1870559} (\bibinfo {year} {2021})}\BibitemShut
  {NoStop}%
\bibitem [{\citenamefont {White}\ \emph {et~al.}(2020)\citenamefont {White},
  \citenamefont {Hill}, \citenamefont {Pollock}, \citenamefont {Hollenberg},\
  and\ \citenamefont {Modi}}]{White2020}%
  \BibitemOpen
  \bibfield  {author} {\bibinfo {author} {\bibfnamefont {G.~A.~L.}\
  \bibnamefont {White}}, \bibinfo {author} {\bibfnamefont {C.~D.}\ \bibnamefont
  {Hill}}, \bibinfo {author} {\bibfnamefont {F.~A.}\ \bibnamefont {Pollock}},
  \bibinfo {author} {\bibfnamefont {L.~C.~L.}\ \bibnamefont {Hollenberg}},\
  and\ \bibinfo {author} {\bibfnamefont {K.}~\bibnamefont {Modi}},\ }\bibfield
  {title} {\bibinfo {title} {Demonstration of non-{M}arkovian process
  characterisation and control on a quantum processor},\ }\href
  {https://doi.org/10.1038/s41467-020-20113-3} {\bibfield  {journal} {\bibinfo
  {journal} {Nature Communications}\ }\textbf {\bibinfo {volume} {11}},\
  \bibinfo {pages} {6301} (\bibinfo {year} {2020})}\BibitemShut {NoStop}%
\bibitem [{\citenamefont {Goswami}\ \emph {et~al.}(2021)\citenamefont
  {Goswami}, \citenamefont {Giarmatzi}, \citenamefont {Monterola},
  \citenamefont {Shrapnel}, \citenamefont {Romero},\ and\ \citenamefont
  {Costa}}]{PhysRevA.104.022432}%
  \BibitemOpen
  \bibfield  {author} {\bibinfo {author} {\bibfnamefont {K.}~\bibnamefont
  {Goswami}}, \bibinfo {author} {\bibfnamefont {C.}~\bibnamefont {Giarmatzi}},
  \bibinfo {author} {\bibfnamefont {C.}~\bibnamefont {Monterola}}, \bibinfo
  {author} {\bibfnamefont {S.}~\bibnamefont {Shrapnel}}, \bibinfo {author}
  {\bibfnamefont {J.}~\bibnamefont {Romero}},\ and\ \bibinfo {author}
  {\bibfnamefont {F.}~\bibnamefont {Costa}},\ }\bibfield  {title} {\bibinfo
  {title} {Experimental characterization of a non-{M}arkovian quantum
  process},\ }\href {https://doi.org/10.1103/PhysRevA.104.022432} {\bibfield
  {journal} {\bibinfo  {journal} {Phys. Rev. A}\ }\textbf {\bibinfo {volume}
  {104}},\ \bibinfo {pages} {022432} (\bibinfo {year} {2021})}\BibitemShut
  {NoStop}%
\bibitem [{\citenamefont {Guo}\ \emph {et~al.}(2021)\citenamefont {Guo},
  \citenamefont {Taranto}, \citenamefont {Liu}, \citenamefont {Hu},
  \citenamefont {Huang}, \citenamefont {Li},\ and\ \citenamefont
  {Guo}}]{PhysRevLett.126.230401}%
  \BibitemOpen
  \bibfield  {author} {\bibinfo {author} {\bibfnamefont {Y.}~\bibnamefont
  {Guo}}, \bibinfo {author} {\bibfnamefont {P.}~\bibnamefont {Taranto}},
  \bibinfo {author} {\bibfnamefont {B.-H.}\ \bibnamefont {Liu}}, \bibinfo
  {author} {\bibfnamefont {X.-M.}\ \bibnamefont {Hu}}, \bibinfo {author}
  {\bibfnamefont {Y.-F.}\ \bibnamefont {Huang}}, \bibinfo {author}
  {\bibfnamefont {C.-F.}\ \bibnamefont {Li}},\ and\ \bibinfo {author}
  {\bibfnamefont {G.-C.}\ \bibnamefont {Guo}},\ }\bibfield  {title} {\bibinfo
  {title} {Experimental demonstration of instrument-specific quantum memory
  effects and non-{M}arkovian process recovery for common-cause processes},\
  }\href {https://doi.org/10.1103/PhysRevLett.126.230401} {\bibfield  {journal}
  {\bibinfo  {journal} {Phys. Rev. Lett.}\ }\textbf {\bibinfo {volume} {126}},\
  \bibinfo {pages} {230401} (\bibinfo {year} {2021})}\BibitemShut {NoStop}%
\bibitem [{\citenamefont {Yu}\ \emph {et~al.}(2018)\citenamefont {Yu},
  \citenamefont {Wang}, \citenamefont {Ke}, \citenamefont {Liu}, \citenamefont
  {Meng}, \citenamefont {Li}, \citenamefont {Zhang}, \citenamefont {Chen},
  \citenamefont {Tang}, \citenamefont {Li},\ and\ \citenamefont
  {Guo}}]{yu2018experimental}%
  \BibitemOpen
  \bibfield  {author} {\bibinfo {author} {\bibfnamefont {S.}~\bibnamefont
  {Yu}}, \bibinfo {author} {\bibfnamefont {Y.-T.}\ \bibnamefont {Wang}},
  \bibinfo {author} {\bibfnamefont {Z.-J.}\ \bibnamefont {Ke}}, \bibinfo
  {author} {\bibfnamefont {W.}~\bibnamefont {Liu}}, \bibinfo {author}
  {\bibfnamefont {Y.}~\bibnamefont {Meng}}, \bibinfo {author} {\bibfnamefont
  {Z.-P.}\ \bibnamefont {Li}}, \bibinfo {author} {\bibfnamefont {W.-H.}\
  \bibnamefont {Zhang}}, \bibinfo {author} {\bibfnamefont {G.}~\bibnamefont
  {Chen}}, \bibinfo {author} {\bibfnamefont {J.-S.}\ \bibnamefont {Tang}},
  \bibinfo {author} {\bibfnamefont {C.-F.}\ \bibnamefont {Li}},\ and\ \bibinfo
  {author} {\bibfnamefont {G.-C.}\ \bibnamefont {Guo}},\ }\bibfield  {title}
  {\bibinfo {title} {Experimental investigation of spectra of dynamical maps
  and their relation to non-{M}arkovianity},\ }\href
  {https://doi.org/10.1103/PhysRevLett.120.060406} {\bibfield  {journal}
  {\bibinfo  {journal} {Phys. Rev. Lett.}\ }\textbf {\bibinfo {volume} {120}},\
  \bibinfo {pages} {060406} (\bibinfo {year} {2018})}\BibitemShut {NoStop}%
\bibitem [{\citenamefont {Lu}\ \emph {et~al.}(2020)\citenamefont {Lu},
  \citenamefont {Zhang}, \citenamefont {Liu}, \citenamefont {Nori},
  \citenamefont {Fan},\ and\ \citenamefont {Pan}}]{lu2020observing}%
  \BibitemOpen
  \bibfield  {author} {\bibinfo {author} {\bibfnamefont {Y.-N.}\ \bibnamefont
  {Lu}}, \bibinfo {author} {\bibfnamefont {Y.-R.}\ \bibnamefont {Zhang}},
  \bibinfo {author} {\bibfnamefont {G.-Q.}\ \bibnamefont {Liu}}, \bibinfo
  {author} {\bibfnamefont {F.}~\bibnamefont {Nori}}, \bibinfo {author}
  {\bibfnamefont {H.}~\bibnamefont {Fan}},\ and\ \bibinfo {author}
  {\bibfnamefont {X.-Y.}\ \bibnamefont {Pan}},\ }\bibfield  {title} {\bibinfo
  {title} {Observing information backflow from controllable non-{M}arkovian
  multichannels in diamond},\ }\href
  {https://doi.org/10.1103/PhysRevLett.124.210502} {\bibfield  {journal}
  {\bibinfo  {journal} {Phys. Rev. Lett.}\ }\textbf {\bibinfo {volume} {124}},\
  \bibinfo {pages} {210502} (\bibinfo {year} {2020})}\BibitemShut {NoStop}%
\bibitem [{\citenamefont {Liu}\ and\ \citenamefont
  {Houck}(2017)}]{liu2017quantum}%
  \BibitemOpen
  \bibfield  {author} {\bibinfo {author} {\bibfnamefont {Y.}~\bibnamefont
  {Liu}}\ and\ \bibinfo {author} {\bibfnamefont {A.~A.}\ \bibnamefont
  {Houck}},\ }\bibfield  {title} {\bibinfo {title} {Quantum electrodynamics
  near a photonic bandgap},\ }\href {https://doi.org/10.1038/nphys3834}
  {\bibfield  {journal} {\bibinfo  {journal} {Nat. Phys.}\ }\textbf {\bibinfo
  {volume} {13}},\ \bibinfo {pages} {48} (\bibinfo {year} {2017})}\BibitemShut
  {NoStop}%
\bibitem [{\citenamefont {Krinner}\ \emph {et~al.}(2018)\citenamefont
  {Krinner}, \citenamefont {Stewart}, \citenamefont {Pazmino}, \citenamefont
  {Kwon},\ and\ \citenamefont {Schneble}}]{krinner2018spontaneous}%
  \BibitemOpen
  \bibfield  {author} {\bibinfo {author} {\bibfnamefont {L.}~\bibnamefont
  {Krinner}}, \bibinfo {author} {\bibfnamefont {M.}~\bibnamefont {Stewart}},
  \bibinfo {author} {\bibfnamefont {A.}~\bibnamefont {Pazmino}}, \bibinfo
  {author} {\bibfnamefont {J.}~\bibnamefont {Kwon}},\ and\ \bibinfo {author}
  {\bibfnamefont {D.}~\bibnamefont {Schneble}},\ }\bibfield  {title} {\bibinfo
  {title} {Spontaneous emission of matter waves from a tunable open quantum
  system},\ }\href {https://doi.org/10.1038/s41586-018-0348-z} {\bibfield
  {journal} {\bibinfo  {journal} {Nature (London)}\ }\textbf {\bibinfo {volume}
  {559}},\ \bibinfo {pages} {589} (\bibinfo {year} {2018})}\BibitemShut
  {NoStop}%
\bibitem [{\citenamefont {Kwon}\ \emph {et~al.}(2022)\citenamefont {Kwon},
  \citenamefont {Kim}, \citenamefont {Lanuza},\ and\ \citenamefont
  {Schneble}}]{Kwon2022}%
  \BibitemOpen
  \bibfield  {author} {\bibinfo {author} {\bibfnamefont {J.}~\bibnamefont
  {Kwon}}, \bibinfo {author} {\bibfnamefont {Y.}~\bibnamefont {Kim}}, \bibinfo
  {author} {\bibfnamefont {A.}~\bibnamefont {Lanuza}},\ and\ \bibinfo {author}
  {\bibfnamefont {D.}~\bibnamefont {Schneble}},\ }\bibfield  {title} {\bibinfo
  {title} {Formation of matter-wave polaritons in an optical lattice},\ }\href
  {https://doi.org/10.1038/s41567-022-01565-4} {\bibfield  {journal} {\bibinfo
  {journal} {Nature Physics}\ }\textbf {\bibinfo {volume} {18}},\ \bibinfo
  {pages} {657} (\bibinfo {year} {2022})}\BibitemShut {NoStop}%
\bibitem [{\citenamefont {Bai}\ \emph {et~al.}(2019)\citenamefont {Bai},
  \citenamefont {Peng}, \citenamefont {Luo},\ and\ \citenamefont
  {An}}]{PhysRevLett.123.040402}%
  \BibitemOpen
  \bibfield  {author} {\bibinfo {author} {\bibfnamefont {K.}~\bibnamefont
  {Bai}}, \bibinfo {author} {\bibfnamefont {Z.}~\bibnamefont {Peng}}, \bibinfo
  {author} {\bibfnamefont {H.-G.}\ \bibnamefont {Luo}},\ and\ \bibinfo {author}
  {\bibfnamefont {J.-H.}\ \bibnamefont {An}},\ }\bibfield  {title} {\bibinfo
  {title} {Retrieving ideal precision in noisy quantum optical metrology},\
  }\href {https://doi.org/10.1103/PhysRevLett.123.040402} {\bibfield  {journal}
  {\bibinfo  {journal} {Phys. Rev. Lett.}\ }\textbf {\bibinfo {volume} {123}},\
  \bibinfo {pages} {040402} (\bibinfo {year} {2019})}\BibitemShut {NoStop}%
\bibitem [{\citenamefont {Camati}\ \emph {et~al.}(2020)\citenamefont {Camati},
  \citenamefont {Santos},\ and\ \citenamefont {Serra}}]{PhysRevA.102.012217}%
  \BibitemOpen
  \bibfield  {author} {\bibinfo {author} {\bibfnamefont {P.~A.}\ \bibnamefont
  {Camati}}, \bibinfo {author} {\bibfnamefont {J.~F.~G.}\ \bibnamefont
  {Santos}},\ and\ \bibinfo {author} {\bibfnamefont {R.~M.}\ \bibnamefont
  {Serra}},\ }\bibfield  {title} {\bibinfo {title} {Employing non-{M}arkovian
  effects to improve the performance of a quantum otto refrigerator},\ }\href
  {https://doi.org/10.1103/PhysRevA.102.012217} {\bibfield  {journal} {\bibinfo
   {journal} {Phys. Rev. A}\ }\textbf {\bibinfo {volume} {102}},\ \bibinfo
  {pages} {012217} (\bibinfo {year} {2020})}\BibitemShut {NoStop}%
\bibitem [{\citenamefont {Luchnikov}\ \emph {et~al.}(2020)\citenamefont
  {Luchnikov}, \citenamefont {Vintskevich}, \citenamefont {Grigoriev},\ and\
  \citenamefont {Filippov}}]{PhysRevLett.124.140502}%
  \BibitemOpen
  \bibfield  {author} {\bibinfo {author} {\bibfnamefont {I.~A.}\ \bibnamefont
  {Luchnikov}}, \bibinfo {author} {\bibfnamefont {S.~V.}\ \bibnamefont
  {Vintskevich}}, \bibinfo {author} {\bibfnamefont {D.~A.}\ \bibnamefont
  {Grigoriev}},\ and\ \bibinfo {author} {\bibfnamefont {S.~N.}\ \bibnamefont
  {Filippov}},\ }\bibfield  {title} {\bibinfo {title} {Machine learning
  non-{M}arkovian quantum dynamics},\ }\href
  {https://doi.org/10.1103/PhysRevLett.124.140502} {\bibfield  {journal}
  {\bibinfo  {journal} {Phys. Rev. Lett.}\ }\textbf {\bibinfo {volume} {124}},\
  \bibinfo {pages} {140502} (\bibinfo {year} {2020})}\BibitemShut {NoStop}%
\bibitem [{\citenamefont {Taranto}\ \emph {et~al.}(2021)\citenamefont
  {Taranto}, \citenamefont {Pollock},\ and\ \citenamefont
  {Modi}}]{Taranto2021}%
  \BibitemOpen
  \bibfield  {author} {\bibinfo {author} {\bibfnamefont {P.}~\bibnamefont
  {Taranto}}, \bibinfo {author} {\bibfnamefont {F.~A.}\ \bibnamefont
  {Pollock}},\ and\ \bibinfo {author} {\bibfnamefont {K.}~\bibnamefont
  {Modi}},\ }\bibfield  {title} {\bibinfo {title} {Non-{M}arkovian memory
  strength bounds quantum process recoverability},\ }\href
  {https://doi.org/10.1038/s41534-021-00481-4} {\bibfield  {journal} {\bibinfo
  {journal} {npj Quantum Information}\ }\textbf {\bibinfo {volume} {7}},\
  \bibinfo {pages} {149} (\bibinfo {year} {2021})}\BibitemShut {NoStop}%
\bibitem [{\citenamefont {Bai}\ and\ \citenamefont
  {An}(2023)}]{PhysRevLett.131.050801}%
  \BibitemOpen
  \bibfield  {author} {\bibinfo {author} {\bibfnamefont {S.-Y.}\ \bibnamefont
  {Bai}}\ and\ \bibinfo {author} {\bibfnamefont {J.-H.}\ \bibnamefont {An}},\
  }\bibfield  {title} {\bibinfo {title} {Floquet engineering to overcome no-go
  theorem of noisy quantum metrology},\ }\href
  {https://doi.org/10.1103/PhysRevLett.131.050801} {\bibfield  {journal}
  {\bibinfo  {journal} {Phys. Rev. Lett.}\ }\textbf {\bibinfo {volume} {131}},\
  \bibinfo {pages} {050801} (\bibinfo {year} {2023})}\BibitemShut {NoStop}%
\bibitem [{\citenamefont {Song}\ \emph {et~al.}(2024)\citenamefont {Song},
  \citenamefont {Liu}, \citenamefont {Zhou}, \citenamefont {Yang},\ and\
  \citenamefont {An}}]{PhysRevLett.132.090401}%
  \BibitemOpen
  \bibfield  {author} {\bibinfo {author} {\bibfnamefont {W.-L.}\ \bibnamefont
  {Song}}, \bibinfo {author} {\bibfnamefont {H.-B.}\ \bibnamefont {Liu}},
  \bibinfo {author} {\bibfnamefont {B.}~\bibnamefont {Zhou}}, \bibinfo {author}
  {\bibfnamefont {W.-L.}\ \bibnamefont {Yang}},\ and\ \bibinfo {author}
  {\bibfnamefont {J.-H.}\ \bibnamefont {An}},\ }\bibfield  {title} {\bibinfo
  {title} {Remote charging and degradation suppression for the quantum
  battery},\ }\href {https://doi.org/10.1103/PhysRevLett.132.090401} {\bibfield
   {journal} {\bibinfo  {journal} {Phys. Rev. Lett.}\ }\textbf {\bibinfo
  {volume} {132}},\ \bibinfo {pages} {090401} (\bibinfo {year}
  {2024})}\BibitemShut {NoStop}%
\bibitem [{\citenamefont {Yeo}\ \emph {et~al.}(2010)\citenamefont {Yeo},
  \citenamefont {An},\ and\ \citenamefont {Oh}}]{PhysRevA.82.032340}%
  \BibitemOpen
  \bibfield  {author} {\bibinfo {author} {\bibfnamefont {Y.}~\bibnamefont
  {Yeo}}, \bibinfo {author} {\bibfnamefont {J.-H.}\ \bibnamefont {An}},\ and\
  \bibinfo {author} {\bibfnamefont {C.~H.}\ \bibnamefont {Oh}},\ }\bibfield
  {title} {\bibinfo {title} {Non-{M}arkovian effects on quantum-communication
  protocols},\ }\href {https://doi.org/10.1103/PhysRevA.82.032340} {\bibfield
  {journal} {\bibinfo  {journal} {Phys. Rev. A}\ }\textbf {\bibinfo {volume}
  {82}},\ \bibinfo {pages} {032340} (\bibinfo {year} {2010})}\BibitemShut
  {NoStop}%
\bibitem [{\citenamefont {Wang}\ \emph {et~al.}(2023)\citenamefont {Wang},
  \citenamefont {Xue}, \citenamefont {Song},\ and\ \citenamefont
  {Jiang}}]{PhysRevA.108.062406}%
  \BibitemOpen
  \bibfield  {author} {\bibinfo {author} {\bibfnamefont {Y.}~\bibnamefont
  {Wang}}, \bibinfo {author} {\bibfnamefont {S.}~\bibnamefont {Xue}}, \bibinfo
  {author} {\bibfnamefont {H.}~\bibnamefont {Song}},\ and\ \bibinfo {author}
  {\bibfnamefont {M.}~\bibnamefont {Jiang}},\ }\bibfield  {title} {\bibinfo
  {title} {Robust quantum teleportation via a non-markovian channel},\ }\href
  {https://doi.org/10.1103/PhysRevA.108.062406} {\bibfield  {journal} {\bibinfo
   {journal} {Phys. Rev. A}\ }\textbf {\bibinfo {volume} {108}},\ \bibinfo
  {pages} {062406} (\bibinfo {year} {2023})}\BibitemShut {NoStop}%
\bibitem [{\citenamefont {Liu}\ \emph {et~al.}(2020)\citenamefont {Liu},
  \citenamefont {Sun}, \citenamefont {Liu}, \citenamefont {Li}, \citenamefont
  {Guo}, \citenamefont {Hamedani~Raja}, \citenamefont {Lyyra},\ and\
  \citenamefont {Piilo}}]{PhysRevA.102.062208}%
  \BibitemOpen
  \bibfield  {author} {\bibinfo {author} {\bibfnamefont {Z.-D.}\ \bibnamefont
  {Liu}}, \bibinfo {author} {\bibfnamefont {Y.-N.}\ \bibnamefont {Sun}},
  \bibinfo {author} {\bibfnamefont {B.-H.}\ \bibnamefont {Liu}}, \bibinfo
  {author} {\bibfnamefont {C.-F.}\ \bibnamefont {Li}}, \bibinfo {author}
  {\bibfnamefont {G.-C.}\ \bibnamefont {Guo}}, \bibinfo {author} {\bibfnamefont
  {S.}~\bibnamefont {Hamedani~Raja}}, \bibinfo {author} {\bibfnamefont
  {H.}~\bibnamefont {Lyyra}},\ and\ \bibinfo {author} {\bibfnamefont
  {J.}~\bibnamefont {Piilo}},\ }\bibfield  {title} {\bibinfo {title}
  {Experimental realization of high-fidelity teleportation via a non-markovian
  open quantum system},\ }\href {https://doi.org/10.1103/PhysRevA.102.062208}
  {\bibfield  {journal} {\bibinfo  {journal} {Phys. Rev. A}\ }\textbf {\bibinfo
  {volume} {102}},\ \bibinfo {pages} {062208} (\bibinfo {year}
  {2020})}\BibitemShut {NoStop}%
\bibitem [{\citenamefont {He}\ \emph {et~al.}(2015)\citenamefont {He},
  \citenamefont {Rosales-Z\'arate}, \citenamefont {Adesso},\ and\ \citenamefont
  {Reid}}]{PhysRevLett.115.180502}%
  \BibitemOpen
  \bibfield  {author} {\bibinfo {author} {\bibfnamefont {Q.}~\bibnamefont
  {He}}, \bibinfo {author} {\bibfnamefont {L.}~\bibnamefont
  {Rosales-Z\'arate}}, \bibinfo {author} {\bibfnamefont {G.}~\bibnamefont
  {Adesso}},\ and\ \bibinfo {author} {\bibfnamefont {M.~D.}\ \bibnamefont
  {Reid}},\ }\bibfield  {title} {\bibinfo {title} {Secure continuous variable
  teleportation and {E}instein-{P}odolsky-{R}osen steering},\ }\href
  {https://doi.org/10.1103/PhysRevLett.115.180502} {\bibfield  {journal}
  {\bibinfo  {journal} {Phys. Rev. Lett.}\ }\textbf {\bibinfo {volume} {115}},\
  \bibinfo {pages} {180502} (\bibinfo {year} {2015})}\BibitemShut {NoStop}%
\bibitem [{\citenamefont {L\"utkenhaus}\ \emph {et~al.}(1999)\citenamefont
  {L\"utkenhaus}, \citenamefont {Calsamiglia},\ and\ \citenamefont
  {Suominen}}]{PhysRevA.59.3295}%
  \BibitemOpen
  \bibfield  {author} {\bibinfo {author} {\bibfnamefont {N.}~\bibnamefont
  {L\"utkenhaus}}, \bibinfo {author} {\bibfnamefont {J.}~\bibnamefont
  {Calsamiglia}},\ and\ \bibinfo {author} {\bibfnamefont {K.-A.}\ \bibnamefont
  {Suominen}},\ }\bibfield  {title} {\bibinfo {title} {Bell measurements for
  teleportation},\ }\href {https://doi.org/10.1103/PhysRevA.59.3295} {\bibfield
   {journal} {\bibinfo  {journal} {Phys. Rev. A}\ }\textbf {\bibinfo {volume}
  {59}},\ \bibinfo {pages} {3295} (\bibinfo {year} {1999})}\BibitemShut
  {NoStop}%
\bibitem [{\citenamefont {Weedbrook}\ \emph {et~al.}(2012)\citenamefont
  {Weedbrook}, \citenamefont {Pirandola}, \citenamefont {Garc\'{\i}a-Patr\'on},
  \citenamefont {Cerf}, \citenamefont {Ralph}, \citenamefont {Shapiro},\ and\
  \citenamefont {Lloyd}}]{RevModPhys.84.621}%
  \BibitemOpen
  \bibfield  {author} {\bibinfo {author} {\bibfnamefont {C.}~\bibnamefont
  {Weedbrook}}, \bibinfo {author} {\bibfnamefont {S.}~\bibnamefont
  {Pirandola}}, \bibinfo {author} {\bibfnamefont {R.}~\bibnamefont
  {Garc\'{\i}a-Patr\'on}}, \bibinfo {author} {\bibfnamefont {N.~J.}\
  \bibnamefont {Cerf}}, \bibinfo {author} {\bibfnamefont {T.~C.}\ \bibnamefont
  {Ralph}}, \bibinfo {author} {\bibfnamefont {J.~H.}\ \bibnamefont {Shapiro}},\
  and\ \bibinfo {author} {\bibfnamefont {S.}~\bibnamefont {Lloyd}},\ }\bibfield
   {title} {\bibinfo {title} {Gaussian quantum information},\ }\href
  {https://doi.org/10.1103/RevModPhys.84.621} {\bibfield  {journal} {\bibinfo
  {journal} {Rev. Mod. Phys.}\ }\textbf {\bibinfo {volume} {84}},\ \bibinfo
  {pages} {621} (\bibinfo {year} {2012})}\BibitemShut {NoStop}%
\bibitem [{\citenamefont {Li}\ \emph {et~al.}(2003)\citenamefont {Li},
  \citenamefont {Li}, \citenamefont {Zhang},\ and\ \citenamefont
  {Zhu}}]{LiFu-Li:14}%
  \BibitemOpen
  \bibfield  {author} {\bibinfo {author} {\bibfnamefont {F.-L.}\ \bibnamefont
  {Li}}, \bibinfo {author} {\bibfnamefont {H.-R.}\ \bibnamefont {Li}}, \bibinfo
  {author} {\bibfnamefont {J.-X.}\ \bibnamefont {Zhang}},\ and\ \bibinfo
  {author} {\bibfnamefont {S.-Y.}\ \bibnamefont {Zhu}},\ }\bibfield  {title}
  {\bibinfo {title} {Teleported state and its fidelity in quantum teleportation
  of continuous variables},\ }\href
  {https://cpl.iphy.ac.cn/EN/abstract/article_33017.shtml} {\bibfield
  {journal} {\bibinfo  {journal} {Chinese Physics Letters}\ }\textbf {\bibinfo
  {volume} {20}},\ \bibinfo {eid} {14-17} (\bibinfo {year} {2003})}\BibitemShut
  {NoStop}%
\bibitem [{\citenamefont {Breuer}\ and\ \citenamefont
  {Petruccione}(2007)}]{book_open}%
  \BibitemOpen
  \bibfield  {author} {\bibinfo {author} {\bibfnamefont {H.-P.}\ \bibnamefont
  {Breuer}}\ and\ \bibinfo {author} {\bibfnamefont {F.}~\bibnamefont
  {Petruccione}},\ }\href@noop {} {\emph {\bibinfo {title} {The Theory of Open
  Quantum Systems}}}\ (\bibinfo  {publisher} {Oxford University Press,
  Oxford},\ \bibinfo {year} {2007})\BibitemShut {NoStop}%
\bibitem [{\citenamefont {Chizhov}\ \emph {et~al.}(2002)\citenamefont
  {Chizhov}, \citenamefont {Kn\"oll},\ and\ \citenamefont
  {Welsch}}]{PhysRevA.65.022310}%
  \BibitemOpen
  \bibfield  {author} {\bibinfo {author} {\bibfnamefont {A.~V.}\ \bibnamefont
  {Chizhov}}, \bibinfo {author} {\bibfnamefont {L.}~\bibnamefont {Kn\"oll}},\
  and\ \bibinfo {author} {\bibfnamefont {D.-G.}\ \bibnamefont {Welsch}},\
  }\bibfield  {title} {\bibinfo {title} {Continuous-variable quantum
  teleportation through lossy channels},\ }\href
  {https://doi.org/10.1103/PhysRevA.65.022310} {\bibfield  {journal} {\bibinfo
  {journal} {Phys. Rev. A}\ }\textbf {\bibinfo {volume} {65}},\ \bibinfo
  {pages} {022310} (\bibinfo {year} {2002})}\BibitemShut {NoStop}%
\bibitem [{\citenamefont {He}\ \emph {et~al.}(2022)\citenamefont {He},
  \citenamefont {Malaney},\ and\ \citenamefont
  {Aguinaldo}}]{PhysRevA.105.062407}%
  \BibitemOpen
  \bibfield  {author} {\bibinfo {author} {\bibfnamefont {M.}~\bibnamefont
  {He}}, \bibinfo {author} {\bibfnamefont {R.}~\bibnamefont {Malaney}},\ and\
  \bibinfo {author} {\bibfnamefont {R.}~\bibnamefont {Aguinaldo}},\ }\bibfield
  {title} {\bibinfo {title} {Teleportation of discrete-variable qubits via
  continuous-variable lossy channels},\ }\href
  {https://doi.org/10.1103/PhysRevA.105.062407} {\bibfield  {journal} {\bibinfo
   {journal} {Phys. Rev. A}\ }\textbf {\bibinfo {volume} {105}},\ \bibinfo
  {pages} {062407} (\bibinfo {year} {2022})}\BibitemShut {NoStop}%
\bibitem [{\citenamefont {Cooper}\ \emph {et~al.}(2012)\citenamefont {Cooper},
  \citenamefont {Hallwood}, \citenamefont {Dunningham},\ and\ \citenamefont
  {Brand}}]{PhysRevLett.108.130402}%
  \BibitemOpen
  \bibfield  {author} {\bibinfo {author} {\bibfnamefont {J.~J.}\ \bibnamefont
  {Cooper}}, \bibinfo {author} {\bibfnamefont {D.~W.}\ \bibnamefont
  {Hallwood}}, \bibinfo {author} {\bibfnamefont {J.~A.}\ \bibnamefont
  {Dunningham}},\ and\ \bibinfo {author} {\bibfnamefont {J.}~\bibnamefont
  {Brand}},\ }\bibfield  {title} {\bibinfo {title} {Robust quantum enhanced
  phase estimation in a multimode interferometer},\ }\href
  {https://doi.org/10.1103/PhysRevLett.108.130402} {\bibfield  {journal}
  {\bibinfo  {journal} {Phys. Rev. Lett.}\ }\textbf {\bibinfo {volume} {108}},\
  \bibinfo {pages} {130402} (\bibinfo {year} {2012})}\BibitemShut {NoStop}%
\bibitem [{\citenamefont {Leggett}\ \emph {et~al.}(1987)\citenamefont
  {Leggett}, \citenamefont {Chakravarty}, \citenamefont {Dorsey}, \citenamefont
  {Fisher}, \citenamefont {Garg},\ and\ \citenamefont
  {Zwerger}}]{leggett1987dynamics}%
  \BibitemOpen
  \bibfield  {author} {\bibinfo {author} {\bibfnamefont {A.~J.}\ \bibnamefont
  {Leggett}}, \bibinfo {author} {\bibfnamefont {S.}~\bibnamefont
  {Chakravarty}}, \bibinfo {author} {\bibfnamefont {A.~T.}\ \bibnamefont
  {Dorsey}}, \bibinfo {author} {\bibfnamefont {M.~P.~A.}\ \bibnamefont
  {Fisher}}, \bibinfo {author} {\bibfnamefont {A.}~\bibnamefont {Garg}},\ and\
  \bibinfo {author} {\bibfnamefont {W.}~\bibnamefont {Zwerger}},\ }\bibfield
  {title} {\bibinfo {title} {Dynamics of the dissipative two-state system},\
  }\href {https://doi.org/10.1103/RevModPhys.59.1} {\bibfield  {journal}
  {\bibinfo  {journal} {Rev. Mod. Phys.}\ }\textbf {\bibinfo {volume} {59}},\
  \bibinfo {pages} {1} (\bibinfo {year} {1987})}\BibitemShut {NoStop}%
\bibitem [{\citenamefont {Tong}\ \emph {et~al.}(2010)\citenamefont {Tong},
  \citenamefont {An}, \citenamefont {Luo},\ and\ \citenamefont
  {Oh}}]{PhysRevA.81.052330}%
  \BibitemOpen
  \bibfield  {author} {\bibinfo {author} {\bibfnamefont {Q.-J.}\ \bibnamefont
  {Tong}}, \bibinfo {author} {\bibfnamefont {J.-H.}\ \bibnamefont {An}},
  \bibinfo {author} {\bibfnamefont {H.-G.}\ \bibnamefont {Luo}},\ and\ \bibinfo
  {author} {\bibfnamefont {C.~H.}\ \bibnamefont {Oh}},\ }\bibfield  {title}
  {\bibinfo {title} {Mechanism of entanglement preservation},\ }\href
  {https://doi.org/10.1103/PhysRevA.81.052330} {\bibfield  {journal} {\bibinfo
  {journal} {Phys. Rev. A}\ }\textbf {\bibinfo {volume} {81}},\ \bibinfo
  {pages} {052330} (\bibinfo {year} {2010})}\BibitemShut {NoStop}%
\bibitem [{\citenamefont {An}\ and\ \citenamefont {Zhang}(2007)}]{an2007non1}%
  \BibitemOpen
  \bibfield  {author} {\bibinfo {author} {\bibfnamefont {J.-H.}\ \bibnamefont
  {An}}\ and\ \bibinfo {author} {\bibfnamefont {W.-M.}\ \bibnamefont {Zhang}},\
  }\bibfield  {title} {\bibinfo {title} {Non-{M}arkovian entanglement dynamics
  of noisy continuous-variable quantum channels},\ }\href
  {https://doi.org/10.1103/PhysRevA.76.042127} {\bibfield  {journal} {\bibinfo
  {journal} {Phys. Rev. A}\ }\textbf {\bibinfo {volume} {76}},\ \bibinfo
  {pages} {042127} (\bibinfo {year} {2007})}\BibitemShut {NoStop}%
\bibitem [{\citenamefont {An}\ \emph {et~al.}(2008)\citenamefont {An},
  \citenamefont {Yeo}, \citenamefont {Zhang},\ and\ \citenamefont
  {Oh}}]{An_2009}%
  \BibitemOpen
  \bibfield  {author} {\bibinfo {author} {\bibfnamefont {J.-H.}\ \bibnamefont
  {An}}, \bibinfo {author} {\bibfnamefont {Y.}~\bibnamefont {Yeo}}, \bibinfo
  {author} {\bibfnamefont {W.-M.}\ \bibnamefont {Zhang}},\ and\ \bibinfo
  {author} {\bibfnamefont {C.~H.}\ \bibnamefont {Oh}},\ }\bibfield  {title}
  {\bibinfo {title} {Entanglement oscillation and survival induced by
  non-{M}arkovian decoherence dynamics of the entangled squeezed state},\
  }\href {https://doi.org/10.1088/1751-8113/42/1/015302} {\bibfield  {journal}
  {\bibinfo  {journal} {Journal of Physics A: Mathematical and Theoretical}\
  }\textbf {\bibinfo {volume} {42}},\ \bibinfo {pages} {015302} (\bibinfo
  {year} {2008})}\BibitemShut {NoStop}%
\bibitem [{\citenamefont {Wu}\ \emph {et~al.}(2021)\citenamefont {Wu},
  \citenamefont {Bai},\ and\ \citenamefont {An}}]{wu2021non}%
  \BibitemOpen
  \bibfield  {author} {\bibinfo {author} {\bibfnamefont {W.}~\bibnamefont
  {Wu}}, \bibinfo {author} {\bibfnamefont {S.-Y.}\ \bibnamefont {Bai}},\ and\
  \bibinfo {author} {\bibfnamefont {J.-H.}\ \bibnamefont {An}},\ }\bibfield
  {title} {\bibinfo {title} {Non-{M}arkovian sensing of a quantum reservoir},\
  }\href {https://doi.org/10.1103/PhysRevA.103.L010601} {\bibfield  {journal}
  {\bibinfo  {journal} {Phys. Rev. A}\ }\textbf {\bibinfo {volume} {103}},\
  \bibinfo {pages} {L010601} (\bibinfo {year} {2021})}\BibitemShut {NoStop}%
\bibitem [{\citenamefont {Massar}\ and\ \citenamefont
  {Popescu}(1995)}]{PhysRevLett.74.1259}%
  \BibitemOpen
  \bibfield  {author} {\bibinfo {author} {\bibfnamefont {S.}~\bibnamefont
  {Massar}}\ and\ \bibinfo {author} {\bibfnamefont {S.}~\bibnamefont
  {Popescu}},\ }\bibfield  {title} {\bibinfo {title} {Optimal extraction of
  information from finite quantum ensembles},\ }\href
  {https://doi.org/10.1103/PhysRevLett.74.1259} {\bibfield  {journal} {\bibinfo
   {journal} {Phys. Rev. Lett.}\ }\textbf {\bibinfo {volume} {74}},\ \bibinfo
  {pages} {1259} (\bibinfo {year} {1995})}\BibitemShut {NoStop}%
\bibitem [{\citenamefont {Braunstein}\ \emph {et~al.}(2001)\citenamefont
  {Braunstein}, \citenamefont {Fuchs}, \citenamefont {Kimble},\ and\
  \citenamefont {van Loock}}]{PhysRevA.64.022321}%
  \BibitemOpen
  \bibfield  {author} {\bibinfo {author} {\bibfnamefont {S.~L.}\ \bibnamefont
  {Braunstein}}, \bibinfo {author} {\bibfnamefont {C.~A.}\ \bibnamefont
  {Fuchs}}, \bibinfo {author} {\bibfnamefont {H.~J.}\ \bibnamefont {Kimble}},\
  and\ \bibinfo {author} {\bibfnamefont {P.}~\bibnamefont {van Loock}},\
  }\bibfield  {title} {\bibinfo {title} {Quantum versus classical domains for
  teleportation with continuous variables},\ }\href
  {https://doi.org/10.1103/PhysRevA.64.022321} {\bibfield  {journal} {\bibinfo
  {journal} {Phys. Rev. A}\ }\textbf {\bibinfo {volume} {64}},\ \bibinfo
  {pages} {022321} (\bibinfo {year} {2001})}\BibitemShut {NoStop}%
\bibitem [{\citenamefont {Hammerer}\ \emph {et~al.}(2005)\citenamefont
  {Hammerer}, \citenamefont {Wolf}, \citenamefont {Polzik},\ and\ \citenamefont
  {Cirac}}]{PhysRevLett.94.150503}%
  \BibitemOpen
  \bibfield  {author} {\bibinfo {author} {\bibfnamefont {K.}~\bibnamefont
  {Hammerer}}, \bibinfo {author} {\bibfnamefont {M.~M.}\ \bibnamefont {Wolf}},
  \bibinfo {author} {\bibfnamefont {E.~S.}\ \bibnamefont {Polzik}},\ and\
  \bibinfo {author} {\bibfnamefont {J.~I.}\ \bibnamefont {Cirac}},\ }\bibfield
  {title} {\bibinfo {title} {Quantum benchmark for storage and transmission of
  coherent states},\ }\href {https://doi.org/10.1103/PhysRevLett.94.150503}
  {\bibfield  {journal} {\bibinfo  {journal} {Phys. Rev. Lett.}\ }\textbf
  {\bibinfo {volume} {94}},\ \bibinfo {pages} {150503} (\bibinfo {year}
  {2005})}\BibitemShut {NoStop}%
\bibitem [{\citenamefont {Baur}\ \emph {et~al.}(2012)\citenamefont {Baur},
  \citenamefont {Fedorov}, \citenamefont {Steffen}, \citenamefont {Filipp},
  \citenamefont {da~Silva},\ and\ \citenamefont
  {Wallraff}}]{PhysRevLett.108.040502}%
  \BibitemOpen
  \bibfield  {author} {\bibinfo {author} {\bibfnamefont {M.}~\bibnamefont
  {Baur}}, \bibinfo {author} {\bibfnamefont {A.}~\bibnamefont {Fedorov}},
  \bibinfo {author} {\bibfnamefont {L.}~\bibnamefont {Steffen}}, \bibinfo
  {author} {\bibfnamefont {S.}~\bibnamefont {Filipp}}, \bibinfo {author}
  {\bibfnamefont {M.~P.}\ \bibnamefont {da~Silva}},\ and\ \bibinfo {author}
  {\bibfnamefont {A.}~\bibnamefont {Wallraff}},\ }\bibfield  {title} {\bibinfo
  {title} {Benchmarking a quantum teleportation protocol in superconducting
  circuits using tomography and an entanglement witness},\ }\href
  {https://doi.org/10.1103/PhysRevLett.108.040502} {\bibfield  {journal}
  {\bibinfo  {journal} {Phys. Rev. Lett.}\ }\textbf {\bibinfo {volume} {108}},\
  \bibinfo {pages} {040502} (\bibinfo {year} {2012})}\BibitemShut {NoStop}%
\bibitem [{\citenamefont {Fedorov}\ \emph {et~al.}(2021)\citenamefont
  {Fedorov}, \citenamefont {Renger}, \citenamefont {Pogorzalek}, \citenamefont
  {Candia}, \citenamefont {Chen}, \citenamefont {Nojiri}, \citenamefont
  {Inomata}, \citenamefont {Nakamura}, \citenamefont {Partanen}, \citenamefont
  {Marx}, \citenamefont {Gross},\ and\ \citenamefont
  {Deppe}}]{doi:10.1126/sciadv.abk0891}%
  \BibitemOpen
  \bibfield  {author} {\bibinfo {author} {\bibfnamefont {K.~G.}\ \bibnamefont
  {Fedorov}}, \bibinfo {author} {\bibfnamefont {M.}~\bibnamefont {Renger}},
  \bibinfo {author} {\bibfnamefont {S.}~\bibnamefont {Pogorzalek}}, \bibinfo
  {author} {\bibfnamefont {R.~D.}\ \bibnamefont {Candia}}, \bibinfo {author}
  {\bibfnamefont {Q.}~\bibnamefont {Chen}}, \bibinfo {author} {\bibfnamefont
  {Y.}~\bibnamefont {Nojiri}}, \bibinfo {author} {\bibfnamefont
  {K.}~\bibnamefont {Inomata}}, \bibinfo {author} {\bibfnamefont
  {Y.}~\bibnamefont {Nakamura}}, \bibinfo {author} {\bibfnamefont
  {M.}~\bibnamefont {Partanen}}, \bibinfo {author} {\bibfnamefont
  {A.}~\bibnamefont {Marx}}, \bibinfo {author} {\bibfnamefont {R.}~\bibnamefont
  {Gross}},\ and\ \bibinfo {author} {\bibfnamefont {F.}~\bibnamefont {Deppe}},\
  }\bibfield  {title} {\bibinfo {title} {Experimental quantum teleportation of
  propagating microwaves},\ }\href {https://doi.org/10.1126/sciadv.abk0891}
  {\bibfield  {journal} {\bibinfo  {journal} {Science Advances}\ }\textbf
  {\bibinfo {volume} {7}},\ \bibinfo {pages} {eabk0891} (\bibinfo {year}
  {2021})}\BibitemShut {NoStop}%
\end{thebibliography}%
\end{document}